\documentclass[useamsfonts]{pasj01}
\twocolumn

\usepackage{url}
\usepackage{natbib}
\usepackage{latexsym}
\usepackage{morefloats}
\usepackage{xspace}
\usepackage[usenames, dvipsnames]{xcolor}
\usepackage{colortbl}
\usepackage{multirow}

\DeclareGraphicsExtensions{.pdf,.png,.eps}

\def\amin{$^\prime$}
\def\asec{$^{\prime\prime}$}

\def\lax{{$\mathrel{\hbox{\rlap{\hbox{\lower4pt\hbox{$\sim$}}}\hbox{$<$}}}$}}
\def\gax{{$\mathrel{\hbox{\rlap{\hbox{\lower4pt\hbox{$\sim$}}}\hbox{$>$}}}$}}
\def\simlt{\lower.5ex\hbox{$\; \buildrel < \over \sim \;$}}
\def\simgt{\lower.5ex\hbox{$\; \buildrel > \over \sim \;$}}

\def\sb{mag~arcsec$^{-2}$}

\def\etal{{\ et al.~}}

\def\ser{{S\'{e}rsic\ }}

\def\m2l{{$M_{\star}/L_{\star}$}~}

\def\hscpipe{\texttt{hscPipe}}
\def\synpipe{\texttt{SynPipe}}
\def\cmodel{\texttt{cModel}}
\def\forced{\texttt{forced}}
\def\coadd{\texttt{coadd}}
\def\unforced{\texttt{unforced}}
\def\tract{\texttt{tract}}
\def\visit{\texttt{visit}}
\def\tracts{\texttt{tracts}}
\def\visits{\texttt{visits}}
\def\galsim{\texttt{G}{\scriptsize \texttt{AL}}\texttt{S}{\scriptsize \texttt{IM}}}
\def\hst{{\textit{HST}}}
\def\s2n{{$\mathrm{S}/\mathrm{N}$}}

\newcommand{\term}[1]{\textbf{\texttt{#1}}}

\newcommand{\plus}{\raisebox{.4\height}{\scalebox{.6}{+}}}
\newcommand{\minus}{\raisebox{.4\height}{\scalebox{.8}{-}}}
\newcommand{\smag}{$\sigma_{\Delta\mathrm{Mag}}$}
\newcommand{\mmag}{$\left<{\Delta\mathrm{Mag}}\right>$}
\newcommand{\scolor}{$\sigma_{\Delta\mathrm{Color}}$}
\newcommand{\mcolor}{$\left<{\Delta\mathrm{Color}}\right>$}
\newcommand{\n}{$\minus$}
  
\begin{document}

\SetRunningHead{Huang et al.}{HSC-SSP SynPipe}

\Received{$\langle$April 2017$\rangle$}
\Accepted{$\langle$2017$\rangle$}
\Published{$\langle$2017$\rangle$}


\title{Characterization and Photometric Performance of the Hyper Suprime-Cam 
       Software Pipeline}

\author{Song Huang \altaffilmark{1,2}
        Alexie Leauthaud \altaffilmark{1,2},
        Ryoma Murata \altaffilmark{2, 4},
        James Bosch \altaffilmark{3},
        Paul Price \altaffilmark{3},
        Robert Lupton \altaffilmark{3},
        Rachel Mandelbaum \altaffilmark{5},
        Claire Lackner \altaffilmark{2},
        Steven Bickerton \altaffilmark{2},
        Satoshi Miyazaki \altaffilmark{6,7}, 
        Jean Coupon \altaffilmark{8},
        Masayuki Tanaka \altaffilmark{6}
        }

\altaffiltext{1}{Department of Astronomy and Astrophysics, University of California,
    Santa Cruz, 1156 High Street, Santa Cruz, CA 95064 USA}
    
\altaffiltext{2}{Kavli Institute for the Physics and Mathematics of the
    Universe, The University of Tokyo Institutes for Advanced Study,
    the University of Tokyo (Kavli IPMU, WPI), Kashiwa 277--8583, Japan}

\altaffiltext{3}{Department of Astrophysical Sciences, Princeton University,
    4 Ivy Lane, Princeton, NJ 08544}

\altaffiltext{4}{Department of Physics, University of Tokyo, Tokyo 113-0033, Japan}

\altaffiltext{5}{McWilliams Center for Cosmology, Department of Physics,
    Carnegie Mellon University, Pittsburgh, PA 15213, USA}

\altaffiltext{6}{National Astronomical Observatory of Japan, 2--21--1 Osawa, Mitaka, 
     Tokyo 181--8588, Japan}

\altaffiltext{7}{SOKENDAI (The Graduate University for Advanced Studies), Mitaka,
    Tokyo, 181--8588, Japan}

\altaffiltext{8}{Department of Astronomy, University of Geneva, ch. d'\'Ecogia 16, 
    1290 Versoix, Switzerland}

\email{song.huang@ipmu.jp}



\maketitle

\begin{abstract}

    The Subaru Strategic Program (SSP) is an ambitious multi-band survey
    using the Hyper Suprime-Cam (HSC) on the Subaru telescope. 
    The Wide layer of the SSP is both wide and deep, reaching a detection limit of 
    $i{\sim}26.0$ mag. 
    At these depths, it is challenging to achieve accurate, unbiased, and consistent
    photometry across all five bands. 
    The HSC data are reduced using a pipeline that builds on the prototype pipeline 
    for the Large Synoptic Survey Telescope. 
    We have developed a \texttt{Python}-based, flexible framework to inject synthetic
    galaxies into real HSC images called \synpipe{}. 
    Here we explain the design and implementation of \synpipe{} and generate 
    a sample of synthetic galaxies to examine the photometric performance of the HSC
    pipeline. 
    For stars, we achieve 1\% photometric precision at $i{\sim}19.0$ mag and 6\%
    precision at $i{\sim}25.0$ in the $i$-band (corresponding to statistical scatters 
    of ${\sim}0.01$ and ${\sim}0.06$ mag respectively).  
    For synthetic galaxies with single-\ser{} profiles, forced \cmodel{} photometry
    achieves 13\% photometric precision at $i{\sim}20.0$ mag and 18\% precision at 
    $i{\sim}25.0$ in the $i$-band (corresponding to statistical scatters of 
    ${\sim}0.15$ and ${\sim}0.22$ mag respectively). 
    We show that both \forced{} PSF and \cmodel{} photometry yield unbiased color 
    estimates that are robust to seeing conditions. 
    We identify several caveats that apply to the version of HSC pipeline used for 
    the first public HSC data release (DR1) that need to be taking into consideration. 
    First, the degree to which an object is blended with other objects impacts the 
    overall photometric performance.  
    This is especially true for point sources. 
    Highly blended objects tend to have larger photometric uncertainties, 
    systematically underestimated fluxes and slightly biased colors. 
    Second, $>20$\% of stars at $22.5< i < 25.0$ mag can be misclassified as 
    extended objects.
    Third, the current \cmodel{} algorithm tends to strongly underestimate the 
    half-light radius and ellipticity of galaxy with $i>21.5$ mag.

\end{abstract}


\section{Introduction}
    \label{sec:intro}

    Wide-field, multi-band imaging surveys have stepped onto the central stage of
    modern astrophysics and cosmology over the past decade.  
    These efforts will soon be replaced with even more ambitious programs such as the 
    Large Synoptic Survey Telescope
    (LSST)\footnote{\url{https://www.lsst.org/}}, the Wide-Field Infrared Survey
    Telescope
    (WFIRST; \citealt{Dressler2012, 
    Spergel2015})\footnote{\url{https://wfirst.gsfc.nasa.gov/}},
    and the \textit{Euclid} project
    (\citealt{Laureijs2012})\footnote{\url{http://sci.esa.int/euclid/}}. 
    Among many ongoing efforts, the Subaru Strategic Program
    (SSP; \citealt{HSCDR1})\footnote{\url{http://hsc.mtk.nao.ac.jp/ssp/}}, which
    uses the Hyper Suprime-Cam (HSC; \citealt{Miyazaki2012}) on the prime focus of
    the Subaru telescope, is the most efficient in terms of
    etendue\footnote{The collecting area multiplied by the field of view.}. 
    Surveys such as HSC, and other to follow, will provide stringent constraints on 
    the cosmological model, characterize the evolution of galaxies, map out the 
    stellar structure of our Milky Way, and are poised to discover a large number of
    interesting transient objects.

    Before we can tackle outstanding scientific questions, we must first learn
    how to handle the the large amounts of data\footnote{Until Feb 2017, SSP has 
    accumulated ${\sim}300$ TB of data products.} generated by these projects while 
    satisfying strict requirements for high quality image processing with accurate 
    measurement for the magnitudes and shapes of stars and galaxies. 
    Data handling becomes increasingly challenging in the age of modern imaging
    surveys am we aim to characterize and account for subtle effects related to 
    charge-coupled devices (CCDs). 
    Cameras are made up of multiple CCDs, each with slightly different
    characteristics, and have large fields-of-views (FoV) over focal planes that 
    are not perfectly flat. 
    During observations, the seeing and background conditions display spatial and 
    temporal variations across the FoV. 
    The full-depletion, thick CCDs selected for HSC enable long exposure
    times and have excellent red-sensitivity, but they also suffer from the
    so-called ``brighter-fatter'' effect (\citealt{Antilogus2014, Guyonnet2015}), 
    which means that brighter stars have larger Point Spread Functions (PSFs) than 
    fainter stars. 
    These variations and effects make accurate astrometric and photometric calibration,
    background subtraction, and point spread function (PSF) modeling intrinsically
    difficult (e.g., \citealt{Schlafly2012}). 
    Furthermore, as surveys reach to increasingly deeper detection limits
    (e.g., the SSP Wide layer reaches a $5\ \sigma$ point source detection limit of
    ${\sim}26.4$ mag in $i$-band), the number of objects per unit area increases.  
    Hence, deep surveys become more sensitive and subject to the effects of blending.  
    The challenge of modern imaging surveys call for  photometric measurement methods
    that are more powerful and precise than those used in previous, shallower surveys.
    
    These challenges are not merely technical details; their resolution is crucial
    to achieving key scientific goals.
    For instance, weak gravitational lensing (WL; \citealt{Kaiser1993, Bartelmann2001})
    is a powerful tool for measuring the large-scale distribution of dark matter.
    However, WL measurements critically depend on our ability to measure the shape and 
    photometric redshifts of background galaxies with high precision. 
    Photometric redshifts (e.g., \citealt{Benitez2000, Bolzonella2000, Ilbert2009}) are 
    fundamentally important for studying the evolution of galaxies, and the ``drop-out'' 
    method is critical for selecting high-redshift galaxies 
    (e.g., \citealt{Steidel1996}).
    However, both of those methods rely on the availability of accurate multi-band 
    photometric measurements. 
    With increasingly large surveys, and with more stringent requirements on data 
    quality, quality control also becomes a pressing issue. 
    
    In this paper, we present a \texttt{Python}-based software package called 
    \synpipe{} that injects synthetic objects into HSC images and which interfaces with 
    the HSC data reduction pipeline (\hscpipe{}). 
    Among other applications, \synpipe{} can be used to perform quality control and 
    to characterize the performance and limitations of \hscpipe{}. 
    Several tools with similar goals have been developed (e.g., \citealt{Chang2015,
    Suchyta2016}) for the Dark Energy Survey (DES; \citealt{DES2005}).

    In Section \ref{sec:ssp}, we briefly introduce the HSC survey and the current status
    of data reductionin. 
    In Section \ref{sec:synpipe}, we explain \synpipe{} design and implementation.
    In Section \ref{sec:test}, we demonstrate \synpipe{} usage via straightforward
    tests, we show the main results for the general photometric quality of synthetic 
    stars and galaxies in Section \ref{sec:result}.
    We then summarize the work and discuss future developments in Section 
    \ref{sec:summary}.

    The code for \synpipe{}, along with documentation and examples, is made available
    on GitHub at: \texttt{https://github.com/dr-guangtou/synpipe}.
    

\begin{figure*}
    \begin{center}
        \includegraphics[width=\textwidth]{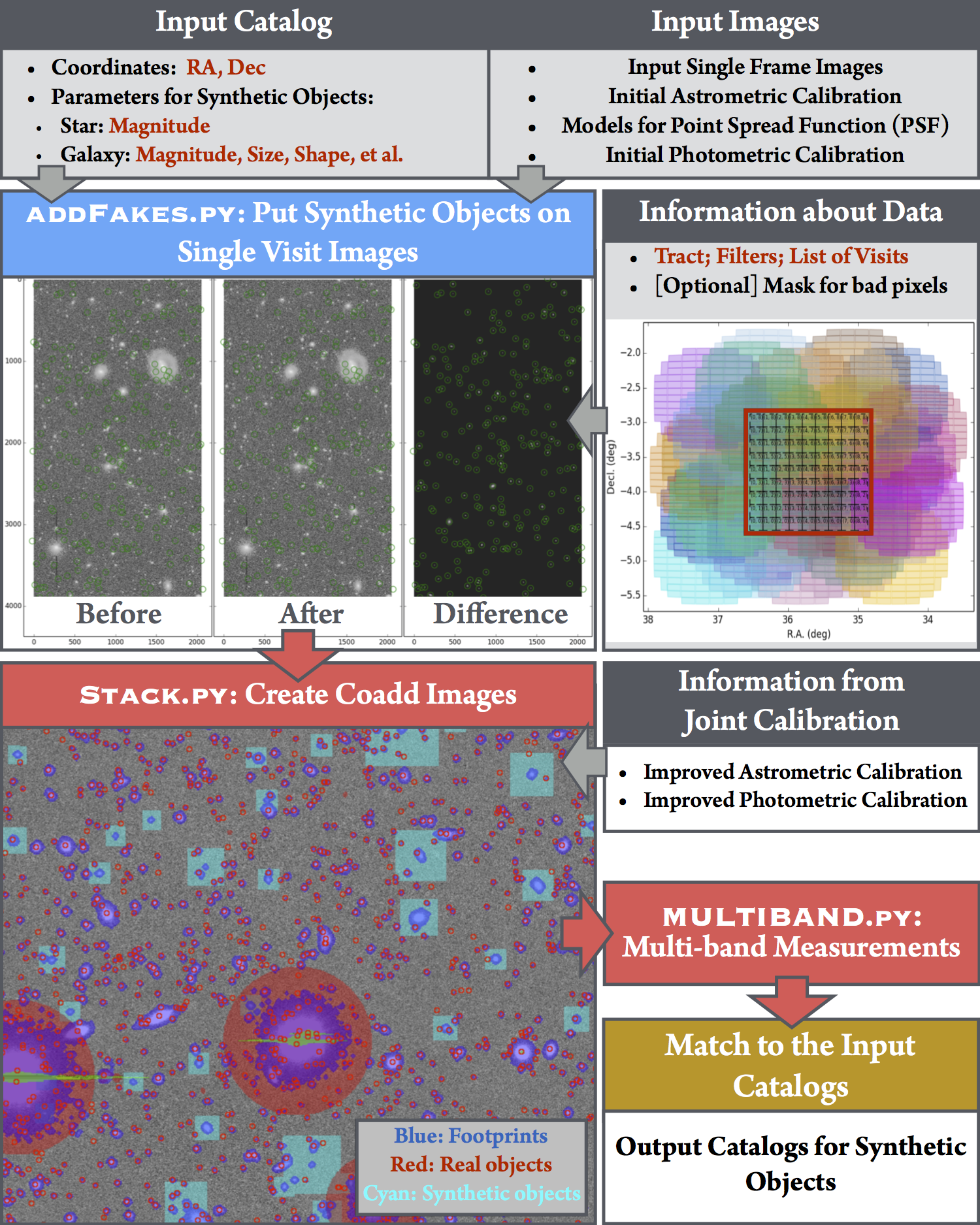}
    \end{center}
    \caption{
        Illustration of the workflow of \synpipe{}.
        Gray boxes indicate required inputs at different stages.
        The blue box identifies the \texttt{addFakes.py} step, when \synpipe{} injects 
        synthetic objects into single-frame images. 
        Below the blue box, we show an HSC image before and after the insertion of 
        synthetic galaxies. 
        The positions of synthetic objects are highlighted with green circles. 
        Red boxes depict the image coadding and multi-band measurement steps
        using \texttt{stack.py} and \texttt{multiBand.py}. 
        At the bottom left, we show a \coadd{} image which contains synthetic galaxies. 
        On the right hand side, we show the spatial relation between \texttt{tracts},
        \texttt{patches}, and \visits{}. 
        The red colored box corresponds to one \tract{} which has an area of 
        about 1.5 deg$^2$. 
        Large colored circles are \visits{} (also commonly known as ``pointings'') 
        with small rectangles representing CCDs. 
        \texttt{patches} are represented by black dashed lines. 
        One \tract{} typically contains 81 \texttt{patches}.
        }
    \label{fig:flowchart}
\end{figure*}

\section{Hyper-Suprime Cam Subaru Strategic Program (HSC Survey)}
    \label{sec:ssp}

\subsection{Status of the Survey}
    \label{ssec:ssp}
    
    Taking advantage of the new prime focus camera on the 8.2--m Subaru telescope,
    the ambitious HSC survey consists of three layers: Wide, Deep, and UltraDeep.
    The Wide layer will map a total of ${\sim}1400$ deg$^2$ of sky in five broad bands
    ($grizy$; Kawanomoto\etal in prep.).
    The Deep (four separated fields; ${\sim}27$ deg$^2$) and UltraDeep (two separated
    fields; ${\sim}3.5$ deg$^2$) layers use a few additional narrow-band filters
    and employ a slightly different surveying strategy.
    \citet{HSCOverview} describes the HSC survey in more detail, and identifies the
    HSC collaborators.

    The HSC camera (\citealt{Miyazaki2012}) is made up of 124 full-depletion thick
    CCDs: 112 for science and another 12 for guiding and focusing.
    The camera has a circular FoV with a 1.5 deg diameter.
    Each CCD contains 2048$\times$4096 pixels and the sizes of pixels are 
    0.168\asec{}.
    For more details about the HSC camera, please see Miyazaki\etal (in prep.).

    In February 2017, the HSC collaboration released the first 1.7 years of data to
    the public (DR1;
    \citealt{HSCDR1})\footnote{\texttt{https://hsc-release.mtk.nao.ac.jp }}.
    For our work, we focused on data from the Wide layer as they relate to key
    scientific goals of the HSC survey, which include weak lensing cosmology, galaxy 
    evolution, and studies of galaxy clusters. 
    The Wide layer data in DR1 correspond to ${\sim}108$ deg$^2$ spread over 
    six fields (XMM--LSS, GAMA09H, WIDE12H, GAMA15H, HECTOMAP, and VVDS)
    \footnote{A small AEGIS field is also observed for calibration purposes.}.
    Except for the HECTOMAP field, the regions are all close to the equator.
    In the Wide layer, the $g$ and $r$ bands have 10 minute exposures broken into 
    four dithers.  
    The $i$, $z$, and $y$ bands have 20 minute exposures broken into six dithers. 
    The survey prioritizes the observations so that the $i$-band has the best seeing 
    conditions to improve galaxy shape measurements for weak lensing science. 
    The $i$ band data in the Wide layer reach a $5\sigma$ point source limiting 
    magnitude of $i{\sim} 26.4$ mag and have a median seeing with 
    FWHM${\sim}0.6$\asec{} in the $i$ band. 
    \citealt{HSCDR1} provides details on the first data release and data status.


\subsection{The HSC Data Reduction Pipeline}
    \label{ssec:hscpipe}

    The HSC data are reduced using a pipeline that builds on the prototype pipeline 
    being designed for the Large Synoptic Survey Telescope’s Data Management system
    (\citealt{Ivezic2008, Axelrod2010, Juric2015}) and is described in 
    Bosch\etal (in prep.). 
    DR1 data are reduced by \hscpipe{} \texttt{v4.0.5}. 
    Because\synpipe{} is intrinsically connected to the complex data reduction
    processes in \hscpipe{}, we briefly introduce the main steps below along with a 
    few key HSC/LSST terms.

    \begin{enumerate}

        \item \textbf{Single-\texttt{Visit~}Processing}:
            Each individual exposure is called a \term{Visit} and is assigned
            an even integer.
            After initial data screening (considering the background level, seeing, 
            and transparency Furusawa\etal in prep.), \hscpipe{} subtracts overscan, 
            bias, and dark frames, and performs flatfielding to the single CCD images.
            During this step, \hscpipe{} also generates variance and mask images, 
            subtracts the background, and provides initial astrometric and photometric
            calibrations. 
            The photometric calibration is based on data from the Panoramic Survey 
            Telescope and Rapid Response System (Pan-STARRS) 1 imaging survey 
            (\citealt{Schlafly2012, Tonry2012, Magnier2013}).
            \synpipe{} uses the spatially varying PSF models, photometric zeropoints, 
            and the World Coordinate System (WCS; \citealt{WCS1, WCS2})
            corrected for optical distortion provided by \hscpipe{}.
            
        \item \textbf{Multi-\visit{} Processing}:
            After single-\visit{} processing, \hscpipe{} warps and mosaics the
            reduced CCD images into much deeper \coadd{} images while improving
            the astrometric and photometric calibrations via processes similar to
            the \"{u}ber-calibration in the Sloan Digital Sky Survey (SDSS)
            \citep{Padmanabhan2008}.
            \synpipe{} will use these improved calibrations.
            \hscpipe{} organizes these \coadd{} images into equiareal rectangular
            regions, or \tracts{}, which are predefined as iso-latitude tessellations.
            One \tract{} covers approximately $1.7\times 1.7$ degrees$^2$ and
            adjacent \tracts{} overlap each other by ${\sim}1$\amin{}.
            Each \tract{} is further divided into $9\times9$ \texttt{Patches}.
            A \texttt{Patch} is a $4200\times4200$ pixels rectangular region, and
            adjacent \texttt{Patches} overlap each other by 100 pixels.   
            
        \item \textbf{Multi-band Measurements}:
            To achieve consistent photometry across all filters, \hscpipe{} first
            detects and deblends objects on \coadd{} images in each band
            independently (the \unforced{} measurements).
            The collection of above-threshold pixels for each object is referred to as
            a \texttt{footprint}.
            \hscpipe{} merges the \texttt{footprints} and flux peaks in the different
            bands and then selects a reference band for each object based on its $S/N$ 
            in each band (usually this corresponds to the $i$-band).  
            Centroids, shapes, and other non-amplitude parameters are then fixed to  
            the values from the reference band. 
            \hscpipe{} then performs \forced{} photometry using these fixed quantities. 
            The goal of the  \forced{} photometry step is to generate accurate colors.
            
        \item \textbf{HSC Photometry}:
            The multi-band catalogs generated by \hscpipe{} contain various types of 
            photometric measurements (see \citealt{HSCDR1} for details). 
            Here, we focus on characterizing the the \forced{} \texttt{psf} and 
            \cmodel{} photometry. 
            The \texttt{psf} photometry should provide the most appropriate magnitudes 
            and colors for point sources, while the \cmodel{} should be used for 
            galaxies. 
            
            \begin{itemize}
            
                \item \hscpipe{} uses matched-filter method to derive PSF magnitude 
                    with the centroid fixed. 
                    The algorithm matches the image of a star with the PSF model 
                    multiplied by an amplitude. 
                    This amplitude parameter is estimated via the inner product of 
                    the image and PSF model after shifting the model to the centroid 
                    of the star, then divide by the effective area of the PSF model.
                    
                \item The HSC \cmodel{} algorithm is described in Bosch\etal (in prep), 
                    and is based on the version for the SDSS \cmodel{} magnitude 
                    (\citealt{Lupton2001, Abazajian2004}). 
                    It fits objects using both a de~Vaucouleurs and an exponential 
                    profiles convolved with the PSF model.  
                    \cmodel{} then linearly combines the two results to find the best 
                    fit to the galaxy image. 
                    \cmodel{} magnitudes are designed to yield accurate fluxes and 
                    colors for galaxies. 
                    Star--galaxy classification is also performed based on the 
                    difference between \texttt{psf} and \cmodel{}{} magnitudes.
                
            \end{itemize}
       
    \end{enumerate}

    DR1 data quality has been vetted in \citet{HSCDR1} by cross-checking with the 
    Pan-STARRS 1 imaging survey to examine the behavior of the stellar sequence.
    Generally speaking, the PSF photometry is accurate at the 1--2\% level for bright 
    stars, and the astrometry is accurate to ${\sim}10$ and 40 max internally and 
    externally.


\section{SynPipe Overview}
    \label{sec:synpipe}

\subsection{Design}
    \label{ssec:design}

    \hscpipe{} is a complex data reduction pipeline that still faces many challenges
    and is under active development.
    Unlike most publicly available photometric pipelines, it involves high-level 
    reduction processes that produce not only deep \coadd{} images but also a series 
    of science-ready catalogs. 
    To perform tests of the data products that result from \hscpipe{}, \synpipe{} is 
    designed to satisfies a few basic standards:

    \begin{itemize}

        \item \textbf{Realistic Images}:
            Instead of using a fully generative approach to simulate full HSC-like
            images from the ground up (e.g., \citealt{Chang2015}), \synpipe{} injects
            synthetic objects into real HSC images so that all the realistic
            features of the real data (e.g., blended objects; proximity to bright
            objects, bleeding trails, other optical artifacts) can be included in the
            test.
            The impact of these features can be important for studies that care about 
            completeness and the selection function.
            In \synpipe{}, point sources are simulated using the HSC PSF model,
            and a realistic galaxy model is created by the modular galaxy image
            simulation toolkit, \galsim{} (\citealt{Rowe2015})
            \footnote{\texttt{https://github.com/GalSim-developers/GalSim}}.

        \item \textbf{Authentic Data Processing}:
            To follow the data reduction process as realistically as possible, we 
            choose to start from the single-\visit{} images instead of directly putting 
            synthetic objects on the final \coadd{} images 
            (e.g., \citealt{Suchyta2016}).
            Through this approach, subtle effects like seeing differences among
            \visits{} and small errors in astrometric calibration can be taken
            into account.
            Every synthetic object will experience all the steps we would use for a 
            real object: detection, deblending, stacking, and measurement.
            This can be important for challenging tasks like WL measurements or the 
            detection of high-$z$ galaxies.
            The downside of this choice is that it slows down the overall run-time
            because we need to inject synthetic objects into all the \visits{} that 
            contribute to a \tract{}.
        
        \item \textbf{Flexible Capabilities}: 
            \synpipe{} is designed to be flexible enough to be useful for HSC users 
            with a range of scientific goals.
            We provide default catalogs to generate samples of synthetic stars
            and galaxies with realistic magnitudes and color distributions along with
            tools to help the user work on HSC DR1 data. 
            However, the users can also supply their own input catalogs best suited 
            for their applications.
            \synpipe{} is already being used in a wide range of
            scientific topics (see Section \ref{sec:summary} for details).

    \end{itemize}

\subsection{\synpipe{} Implementation }
    \label{ssec:flowchart}

    In this section, we describe the \synpipe{} test implementation illustrated in 
    Fig \ref{fig:flowchart}.

\subsubsection{Preparation}
    \label{sssec:prep}

    To create a catalog of synthetic objects, the user first needs to select which HSC 
    data and which input synthetic galaxy catalog to use. 

    \begin{itemize}

        \item \textbf{Information about the data}. 
            This  corresponds to the location of HSC images and a list of \visits{} to 
            be used. 
            \synpipe{} can also help the user identify all the \visits{} that contribute 
            to one \tract{} in any given band.
            \synpipe{} also provides tools to create an optional bad-pixel mask
            (e.g., bright object, bleeding trails) for a specific \tract{} so that
            the user can avoid putting synthetic objects on problematic regions.

        \item \textbf{Input catalog}. 
            This is a catalog in \texttt{FITS} (Flexible Image Transport System)
            format that contains the coordinates, magnitudes, and model parameters of 
            synthetic objects.  
            The positions of synthetic galaxies can be specified via the input catalog, 
            they can be distributed randomly over a given region, or on an 
            evenly-spaced grid (the grid option is useful in that it avoids blends 
            between Synthetic objects, e.g. see Murata\etal in prep.).  
            
    \end{itemize}

    \vspace{0.5cm}
    \noindent Synthetic objects can be one of the following:

    \begin{itemize}

        \item \textbf{Point source}: 
            \synpipe{} simply uses the PSF model from \hscpipe{} as the model for
            a point source (stars and/or quasi-stellar objects).
            \hscpipe{} uses a special version of \texttt{PSFEx}
            (\citealt{Bertin2011, Bertin2013}) to characterize PSF as a function
            of position and the PSF model can be reconstructed for any location on 
            the image. 
            To inject point sources, the user only needs to specify a catalog of 
            coordinates and magnitudes.
            
        \item \textbf{\ser{}models}:
            \texttt{SynPipe} uses \galsim{} \texttt{v1.4} (\citealt{Rowe2015}) to
            simulate galaxies.
            \galsim{} is an image simulation tool that was designed for the 
            GRavitational lEnsing Accuracy Testing 3 (GREAT3) challenge
            (\citealt{Mandelbaum2014}). 
            Currently, \synpipe{} allows synthetic galaxies to be modeled by a
            single or a double \citet{Sersic1963}
            profile\footnote{More flexible model choices will be added later.}. 
            The \ser{} profile is flexible enough to describe the overall flux
            distributions of galaxies near and far, both early-type or late-type.
            A double-\ser{} model can simulate a galaxy with even more realistic
            structural details (e.g., bulge$+$disk).
            For each \ser{} component, the user needs to provide the magnitude,
            effective radius ($R_{\mathrm{e}}$, in unit of arcsec), \ser{} index
            ($n_{\mathrm{Ser}}$), axis ratio ($\mathrm{b}/\mathrm{a}$), and position
            angle (PA).
            External shear can also be applied to a synthetic galaxy ($g_1$ and
            $g_2$, see \citealt{Rowe2015} for more details).
            
        \item \textbf{Galaxies from the GREAT3 challenge}: 
            \synpipe{} also allows users to choose from parametric models or real
            high-resolution galaxy images used in the GREAT3 WL challenge. 
            Real galaxy images are drawn from the \hst{}/ACS F814W ($I$-band) images 
            of galaxies in the COSMOS field (e.g., \citealt{Scoville2007,Leauthaud2007}). 
            Instead of using real galaxy images, the users may also use models of COSMOS 
            galaxies to $I_{\mathrm{Auto}}\leq25.2$ mag via a method developed by
            \citet{Lackner2012}.  
            Parametric models of COSMOS galaxies include a single-\ser{} and the \ser{}
            bulge$+$Exponential disk model. 
            The parameter values for these models are stored in the
            \texttt{GalSim.COSMOSCatalog()} catalog.
            To access them, the user can provide the ID number of COSMOS galaxies
            in the input catalog, or can simply opt to let \synpipe{} randomly select 
            galaxies from the COSMOS input catalog.
            
    \end{itemize}

\begin{figure*}
    \begin{center}
        \includegraphics[width=\textwidth]{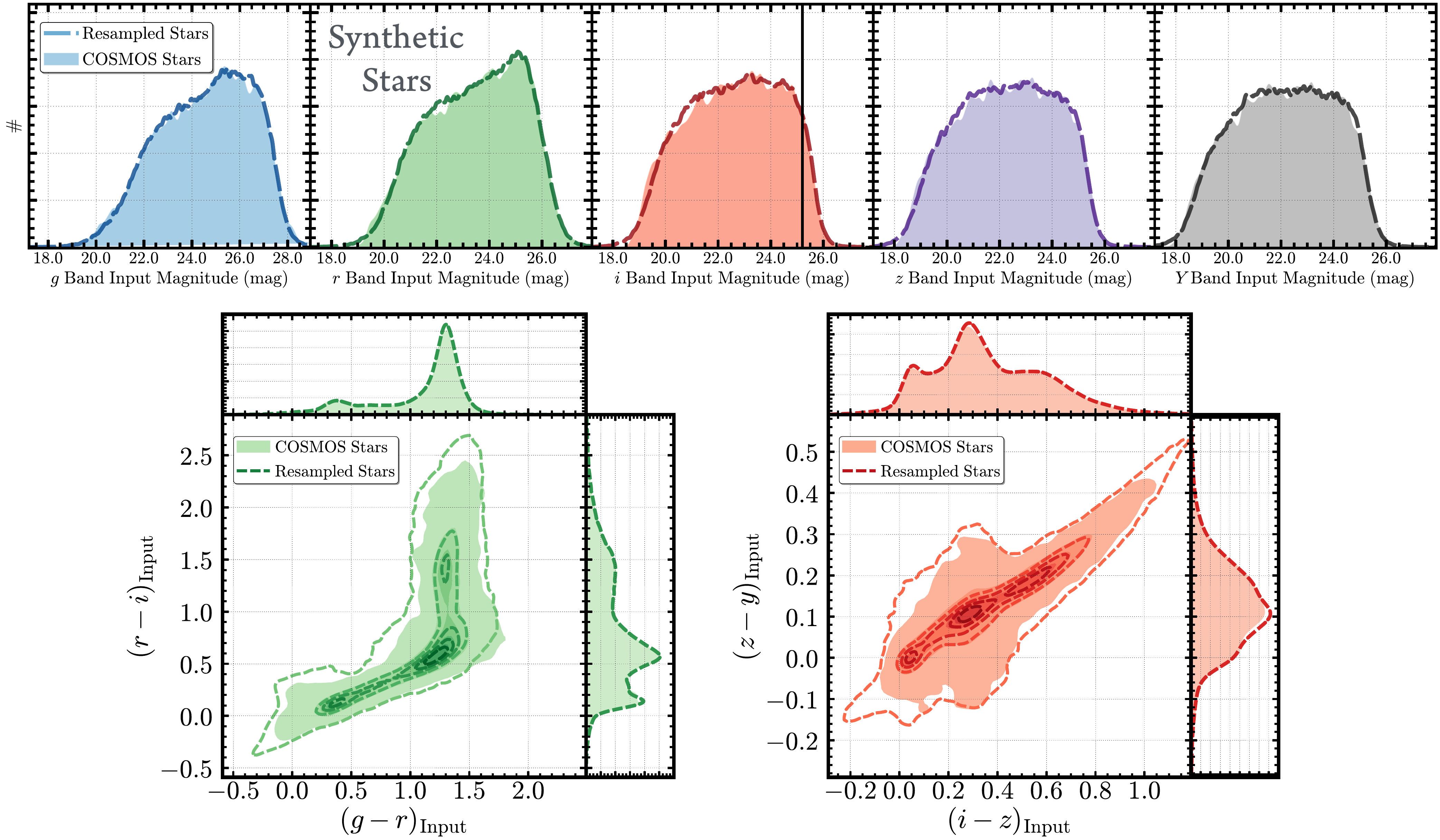}
    \end{center}
    \caption{
        Magnitude and color distributions of synthetic stars.
        \textbf{Upper panel}: five-band magnitude distributions of the COSMOS
        stars (filled histograms) and  resampled stars (dashed lines).
        \textbf{Lower panel}: color-color distributions
        (left: $g-r$ v.s. $r-i$; right: $i-z$ v.s. $z-y$). 
        Filled-contours indicate COSMOS stars and open-contours indicate resampled stars.
        }
    \label{fig:star_sample}
\end{figure*}

\subsubsection{Injection of Synthetic Objects into Single-\visit{} Images}
    \label{sssec:addFakes}
    
    \begin{itemize}
    
        \item With the input catalog and test data information in hand, \synpipe{} 
            injects synthetic objects into single-exposure images (step
            \texttt{addFakes.py}; see the middle panel of Fig \ref{fig:flowchart}).
            \synpipe{} goes through individual CCD images that belong to 
            each \visit{} and decides which synthetic objects from the input catalog 
            need to be injected. 
            \synpipe{} uses the \textit{initial astrometric calibration} of each 
            single \visit{} to convert the input coordinates into locations on the 
            CCD image. 
            For each synthetic object, \synpipe{} uses the reconstructed PSF for each 
            exposure.  
            The photometric zero point from the single-\visit{} calibration is used
            to convert input magnitudes of synthetic objects into fluxes.
    
        \item With the help of \galsim{} module, \synpipe{} simulate the images of
            synthetic objects.
            \begin{itemize}
            
            \item For point sources, \synpipe{} generates a rectangular cutout image 
                of the PSF model with appropriate size and correct total flux. 
            
            \item For galaxies described by single- or double-\ser{} models, 
                \synpipe{} passes the input \ser{} index and effective radius to 
                the \galsim{}\texttt{.Sersic} function to create a \ser{} component.
                After stretching and rotating the galaxy to the expected axis ratio 
                and position angle via the \texttt{shear} and \texttt{rotate} methods,
                \synpipe{} assigns the correct flux to this component.
            
            \item For a double-\ser{} model, \synpipe{} uses the \galsim{}\texttt{.Add}
                method to combine two \ser{} components. 
            
            \item For parametric models from the GREAT3 catalog, \synpipe{} uses the
                \texttt{COSMOSCatalog.makeGalaxy} method to generate models.  
                
            \item To inject real HST galaxy images, \synpipe{} calls the \galsim{}    
                \texttt{.RealGalaxy} method.
            
            \end{itemize}

        If necessary, an additional shear ($g_1$ and $g_2$) can be applied to the 
        model at this point.
        Then, \synpipe{} passes the reconstructed PSF image into a
        \galsim{}\texttt{InterpolatedImage} object and convolves it with the galaxy 
        model using \galsim{}\texttt{.Convolve}.
        After this, \synpipe{} uses the \galsim{}\texttt{.drawImage} method to redner 
        the image of the simulated galaxy.
        A component with a high \ser{} index ($n_{\mathrm{Ser}} > 4$) often requires 
        a large image size to cover all of its flux, so \synpipe{} allows the user 
        to truncate the model at a given radius (N$\times R_{\mathrm{e}}$).
        \synpipe{} will also give warning information when a component with $
        n_{\mathrm{Ser}} > 6$ is encountered because it takes much longer to achieve
        accurate PSF convolution in such a model; and the result is, therefore,
        not a realistic galaxy model.
        
        \item After \synpipe{} creates the image of the simulated galaxy, it then 
            shifts the image according to is location on the CCD on a pixel-by-pixel 
            basis; \synpipe{} also crops the images when necessary.
            When that task is complete, \synpipe{} adds appropriate noise to the 
            image based on the pixels it covers and the calibration information of 
            the detector, and creates a variance image that reflects the influence 
            of the synthetic object.
            Since each HSC CCD consists of 4 amplifiers, each with different 
            characteristics, \synpipe{} carefully creates a corresponding gain map 
            to provide accurate noise level.
            The added noise only accounts for the additional flux in the synthetic 
            object, and will thus be subdominant compared to the noise already in 
            the image for faint objects.
            
        \item In the final step, \synpipe{} adds the noise-added image and the 
            variance map to the original CCD data while also adding a new 
            \texttt{FAKE} mask bit to the mask plane.
            The middle panel of Fig \ref{fig:flowchart} depicts the result of the
            \texttt{addFakes.py} step by showing an example CCD image before and 
            after the synthetic objects are injected.
    
    \end{itemize}

\subsubsection{Stacking and Multi-band Measurements}
    \label{sssec:multiband}

    The newly generated single-\visit{} images have the same format and data
    structure as real HSC data. To perform stacking and measurements, \synpipe{} calls 
    standard \hscpipe{} routines.
    
    The \texttt{stack.py} step takes the improved astrometric and photometric
    calibrations for each \visit{} from the original reduction and creates
    \coadd{} images that contain synthetic objects.
    The \texttt{multiBand.py} step then processes these \coadd{}  images and
    provides standard multi-band measurements in FITS catalogs that are grouped by
    \texttt{Patch} ID.

    Using the infrastructure provided by \hscpipe{}, the user can easily perform these 
    steps in parallel, and the overall efficiency of \synpipe{} is similar to
    the real data reduction process. 
    The data and catalog volumes is also similar to the original data reduction process.

    The final step is to identify Synthetic galaxies in the output catalogs. 
    \synpipe{} can match the input catalog to the output catalog (this contains a mix 
    of real and synthetic galaxies) using a matching radius specified by the user.
    Each unmatched object from the input catalog is also passed to the results catalog
    with a unique label.
    In the case of multiple matches, the user can choose between returning only the
    closest match or returning all objects within the matching radius (the nearest one
    is labeled as such).

\subsection{Limitations and Caveats}
    \label{ssec:caveats}
    
    The limitations of the current \synpipe{} include the following:

    \begin{enumerate}

        \item \citet{HSCDR1} points out that \hscpipe{} tends to over-subtract the
            background around bright objects.
            Unfortunately, \synpipe{} now takes the original background subtraction on
            the single-\visit{} image for granted; hence, it lacks the capability to
            test or help improve this problem.

        \item  \synpipe{} simply adopts the PSF measured by \hscpipe{} and
            uses it as a model of point source and PSF convolution for a galaxy.
            Hence, it cannot be used to test how the uncertainty of the PSF modeling
            affects the photometry and shape measurements.
            This could be important for regions with exquisite seeing 
            (e.g. FWHM${\sim}0.4$\asec{}) that makes the
            PSF modeling difficult, and for accurate WL measurements 
            (see \citealt{HSCDR1}).

        \item \synpipe{} works in a ``unit'' of \visit{} for a single-exposure
            test, and uses a \tract{} to test \coadd{} images.
            In the case of tests that focus on specific regions that are much smaller 
            than the size of a \visit{} or a \tract{} (e.g., quality of photometry
            around rich galaxy clusters), using \synpipe{} often leads to
            low efficiency.

    \end{enumerate}

\begin{figure*}
    \begin{center}
        \includegraphics[width=\textwidth]{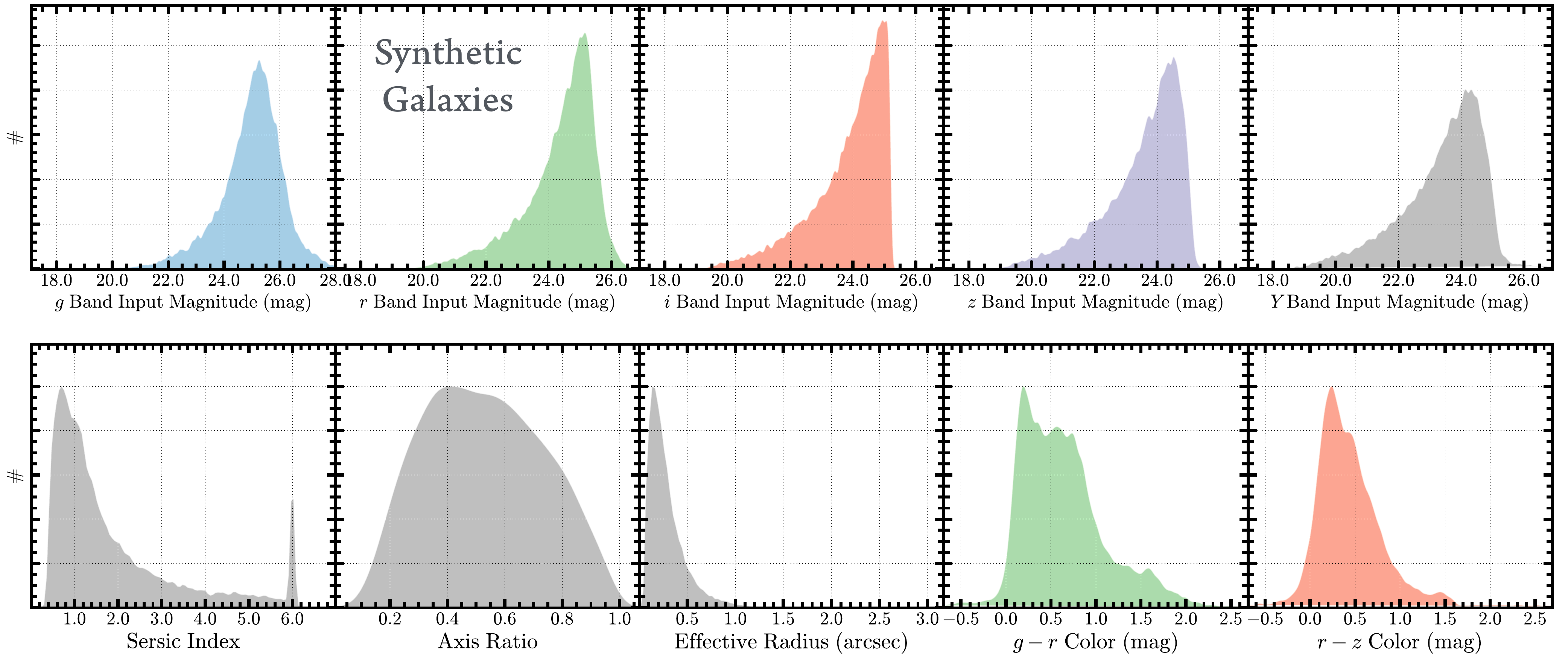}
    \end{center}
    \caption{
         \textbf{Upper panel}: five-band magnitude distributions of synthetic
         galaxies. (\texttt{lower panel}; from left to right are the \ser{} index, 
         axis ratio, effective radius in arcsec, the $g-r$ and $r-z$ colors).
         }
    \label{fig:galaxy_sample}
\end{figure*}

\section{Generation of Synthetic Dataset}
    \label{sec:test}
    
    We now describe the generation of the synthetic data set that we use to 
    characterize the performance of \hscpipe{}.

\subsection{Data}
    We use \tract{}$=9699$ in the VVDS field (median FWHM$=0.449$\asec{} in $i$ band;
    in this paper referred to as \texttt{goodSeeing}) and
    \tract{}$=8764$ in the XMM-LSS field (median FWHM$=0.700$\asec{};
    \texttt{badSeeing}).
    \tract{}$=9699$ also has better seeing in both $r$ and $z$ bands than the median 
    conditions in those bands.
    The seeing conditions in $g$ and $y$ bands are very similar for both
    \tract{}$=9699$ and \tract{}$=8764$.
    For both \tracts{}, the \coadd{} image in each band includes data from
    20--40 \visits{}\footnote{The large number of \visits{} here is due to large 
    dither pattern. Please see Fig \ref{fig:flowchart}.}.  
    These two \tract{} are selected because they are not at the edge of a field, do 
    not contain any extremely bright ($i<12$ mag) saturated stars, and are 
    representative of both "good" and "bad" seeing\footnote{Here, ``bad'' seeing only 
    suggests that it is worse than the median seeing condition by HSC standard.} 
    conditions in the $i$ band.

\begin{figure*}
    \begin{center}
        \includegraphics[width=\textwidth]{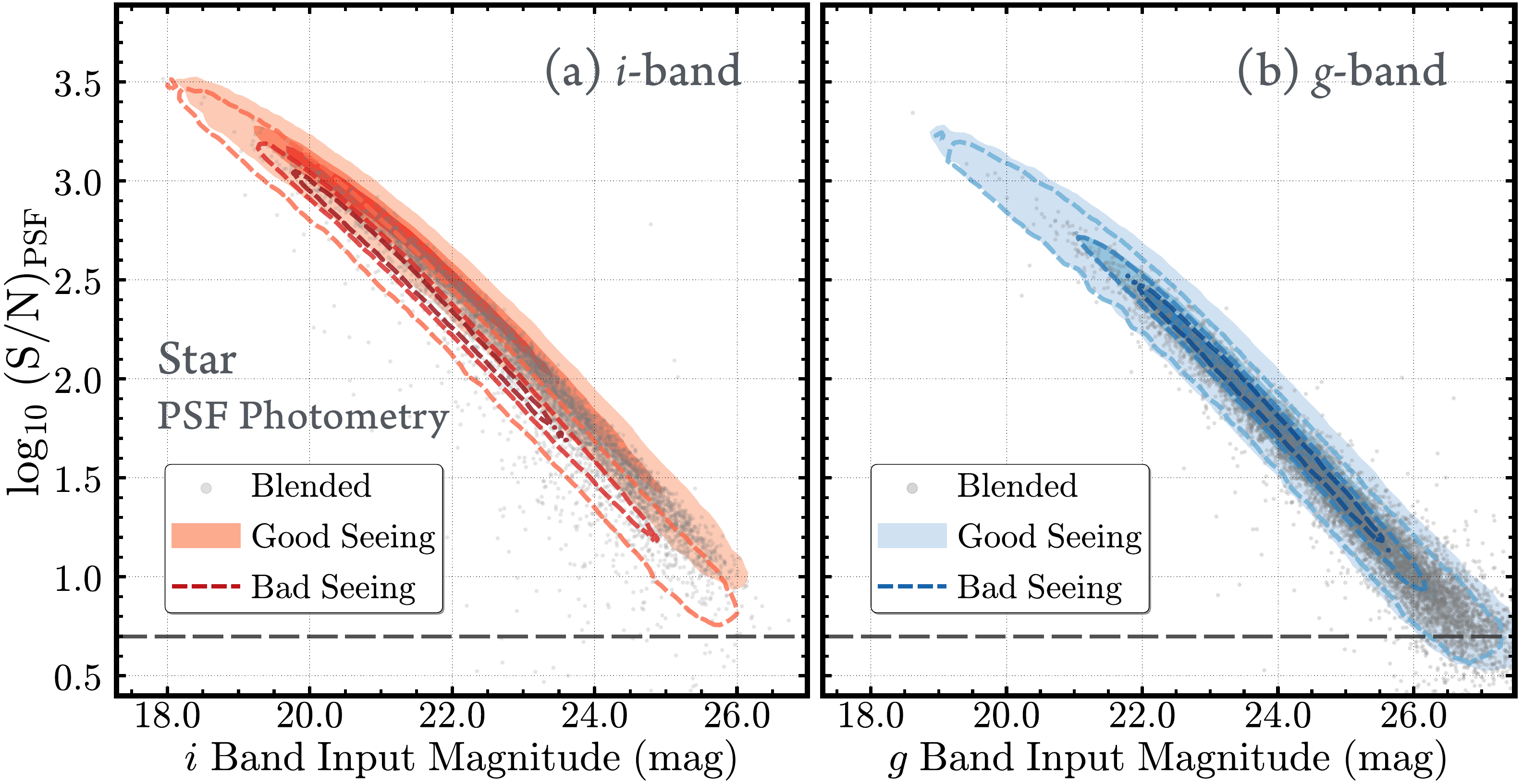}
    \end{center}
    \caption{
        Relation between input magnitudes (\textbf{left}: $i$-band; \textbf{right}:
        $g$-band) of synthetic stars and the $\log (\mathrm{S}/\mathrm{N})$ measured 
        by \hscpipe{} PSF photometry.
        Filled-contours and open-contours show the distributions for
        stars in both the \texttt{goodSeeing} and \texttt{badSeeing} \tracts{}. 
        Highly blended stars are highlighted using scatter plots.
        The gray dashed line marks $S/N = 5$ which is the detection threshold used 
        by \hscpipe{}.  
        Our 5$\sigma$ point source detection limit is ${\sim}26.5$ mag in $i$-band,
        however we do not reach these magnitudes with our current synthetic stellar 
        input catalog.
        }
    \label{fig:star_sn}
\end{figure*}

\subsection{Input Models for Stars and Galaxies}
    \label{ssec:inputs}

    We use a synthetic star sample built using data from the HST/ACS catalog of 
    \citet{Leauthaud2007}.  
    We first select stars from the \citet{Leauthaud2007} catalog (this classification 
    is reliable down to $I_{\mathrm{F814W}}{\sim}25.2$ mag). 
    We then match this sample against our HSC UltraDeep--COSMOS data (which reaches 
    ${\sim}27.2$ mag in $i$ band). 
    After applying basic quality cuts (see \ref{app:qc}), this procedure provides us 
    with five-band HSC PSF photometry for 14,472 stars.
    To increase the sample size, we re-sample the five-band magnitude distributions 
    using the \texttt{astroML} (\citealt{astroml}) implementation of the 
    \texttt{extreme-deconvolution} algorithm developed by \citet{Bovy2011}
    \footnote{\texttt{https://github.com/jobovy/extreme-deconvolution/}} to
    generate 100,000 synthetic stars. 
    Fig \ref{fig:star_sample} shows the magnitude and color distributions of real 
    COSMOS stars compared to the resampled distributions in all five-bands and in 
    color-color space.

    For synthetic galaxies, we use single-\ser{} galaxy models of COSMOS galaxies with 
    $I_{F814W} \leq 25.2$ mag. 
    These are models from \citet{Lackner2012} applied to \textit{HST}/ACS images and 
    are included in \galsim{}. 
    Each model is described by a total of five parameters: magnitude, effective radius,
    \ser{} index, axis ratio, and position angle.
    Although the single-\ser{} model is not always the best choice for describing 
    galaxies, its flexibility enables it to reasonably describe the flux distribution 
    of most galaxies.
    Also, its simplicity makes it easy for us to diagnose potential problems with the 
    \cmodel{} photometry.
    We exclude a tiny fraction (${\sim}4$\%) of ill-behaved models that have a very 
    high \ser{} index ($n_{\mathrm{Ser}} > 6.0$) or a very low central surface 
    brightness ($\mu_{i} < 24.0$\sb).
    We further match this COSMOS sample with the HSC UltraDeep--COSMOS data.
    After removing objects with problematic photometry (see Appendix \ref{app:qc} for 
    details), we obtain a sample of 58,210 synthetic galaxies with realistic 
    five-band HSC \cmodel{} photometry.
    As shown in Fig \ref{fig:galaxy_sample}, the majority of these galaxies are
    faint ($i<24.0$ mag) and barely resolved ($R_{\mathrm{e}}< 1.0$\asec).
    This sample is appropriate to test \hscpipe{}'s general photometric behaviors, 
    but the sample lacks relatively bright galaxies ($i<20.5$ mag).

    We spatially distribute synthetic stars and galaxies randomly in our two selected 
    \tracts{}. 
    For stars, we use a number density of 1,000 per CCD. 
    For galaxies, we use a number density of 500 per CCD. 
    Given the magnitude distributions of these objects, these numbers are high enough 
    to ensure large sample of useful synthetic objects for the test, and will also not
    create unrealistic crowded images.


\subsection{Generating \synpipe{} Data}
    \label{ssec:running}

    We use HSC DR1 data stored at Kavli Institute for the Physics and Mathematics of 
    the Universe (Kavli IPMU) for these tests. 
    Using 108 cores, the \texttt{addFakes.py} step takes ${\sim}1.5$ hours per 
    \tract{} in each band for stars. 
    The same process takes longer for galaxy tests (${\sim}3.0$ hours per
    \tract{} in each band) because the \galsim{} simulation is more time consuming.
    The \texttt{stack.py} step and \texttt{multiBand.py} step together take 
    ${\sim}3.5$ hours per \tract{}.

    We match the results with the input catalogs using a 2--pixel matching radius
    to generate output catalogs for \forced{} and \unforced{} photometry
    in each band. 
    For our tests, we reject synthetic objects locate within 2 pixels 
    of a real HSC object. 
    These ``ambiguously blended'' (e.g. \citealt{Dawson2016}) objects are extreme 
    blends in which multiple objects are detected as a single object, and are not 
    useful for photometric tests.
    
    The detailed log of our runs can be found at \url{goo.gl/VINOVP}.
    The user should be able to reproduce the results presented in this work using 
    this information. It can also serve as a brief manual for generating \synpipe{} 
    data.


\section{Photometric Performance Results}
    \label{sec:result}

    Here we discuss \hscpipe{}'s general performance, assessed mainly through \forced{} 
    PSF photometry (for stars) and (\cmodel{}) photometry (for galaxies).
    Although \hscpipe{} does provide other options (e.g., aperture and Kron photometry),
    these two are the only options that consider the effects of the PSF in different 
    bands and are consistent across all bands in terms of position and shape. 
    

\begin{figure*}
    \begin{center}
        \includegraphics[width=16cm]{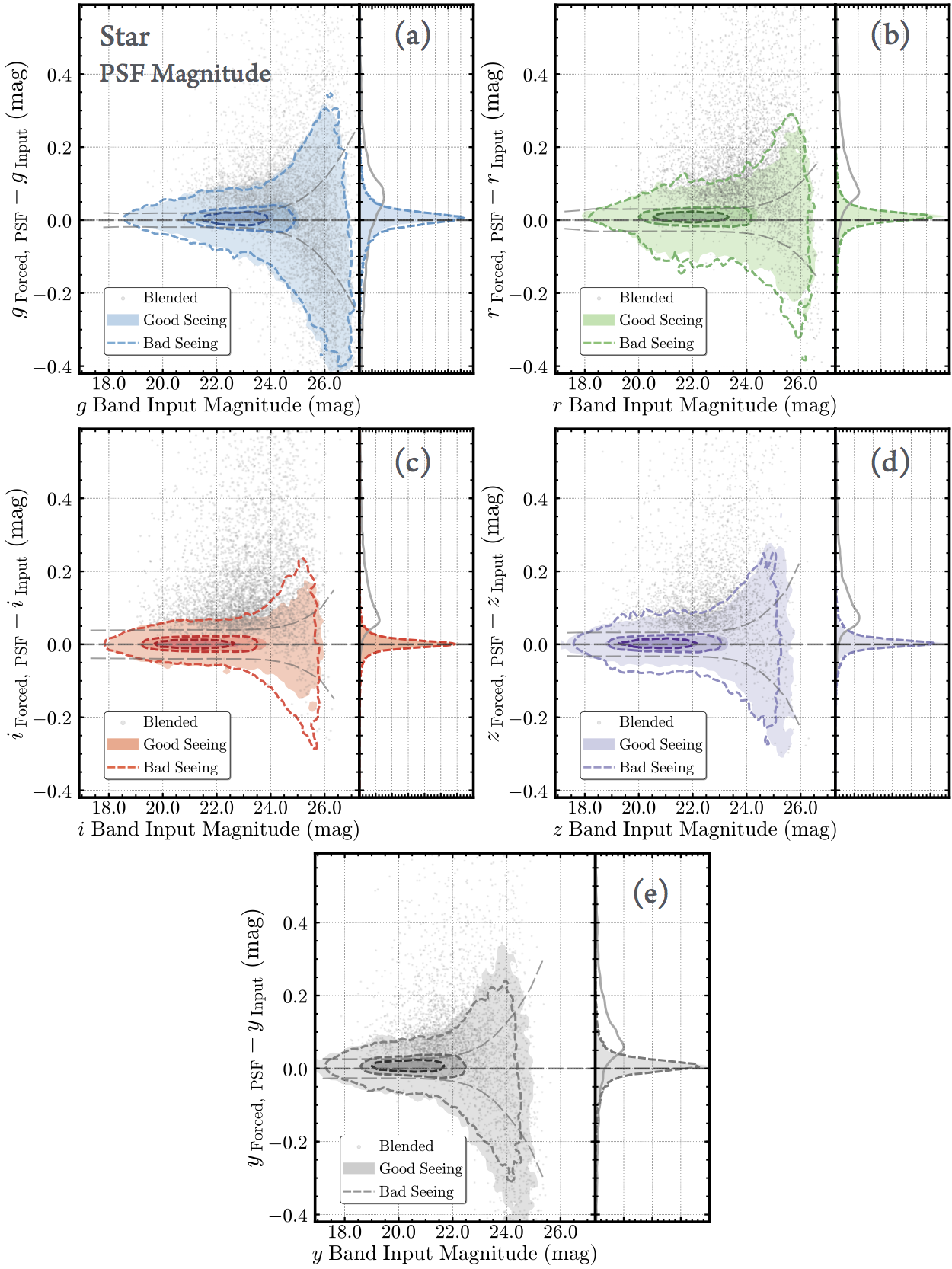}
    \end{center}
    \caption{
        Accuracy of the \texttt{hscpipe} PSF photometry for synthetic stars measured
        by the difference between input and output \forced{} PSF magnitudes.
        Plots [a, b, c, d, e] show the results for [$g$, $r$, $i$, $z$, $y$]-band, 
        respectively.
        The left panel in each plot shows the relation between input magnitude and
        magnitude difference.
        Filled and open contours are for the synthetic galaxies 
        from \texttt{goodSeeing} and \texttt{badSeeing} \tracts{}.
        Highly blended objects are highlighted using scattered points.
        The long-dashed lines mark zero magnitude difference, while the pairs of
        dashed lines outline the running-median of PSF magnitude errors
        (including the uncertainties in aperture correction).
        The right panel in each plot shows the distributions of the magnitude 
        differences for objects in \texttt{goodSeeing} (filled) and \texttt{badSeeing}
        (solid line) \tracts{}.
        The dashed lines identify the distribution of highly blended objects.
        }
    \label{fig:psf_mag}
\end{figure*}

\begin{figure*}
    \begin{center}
        \includegraphics[width=\textwidth]{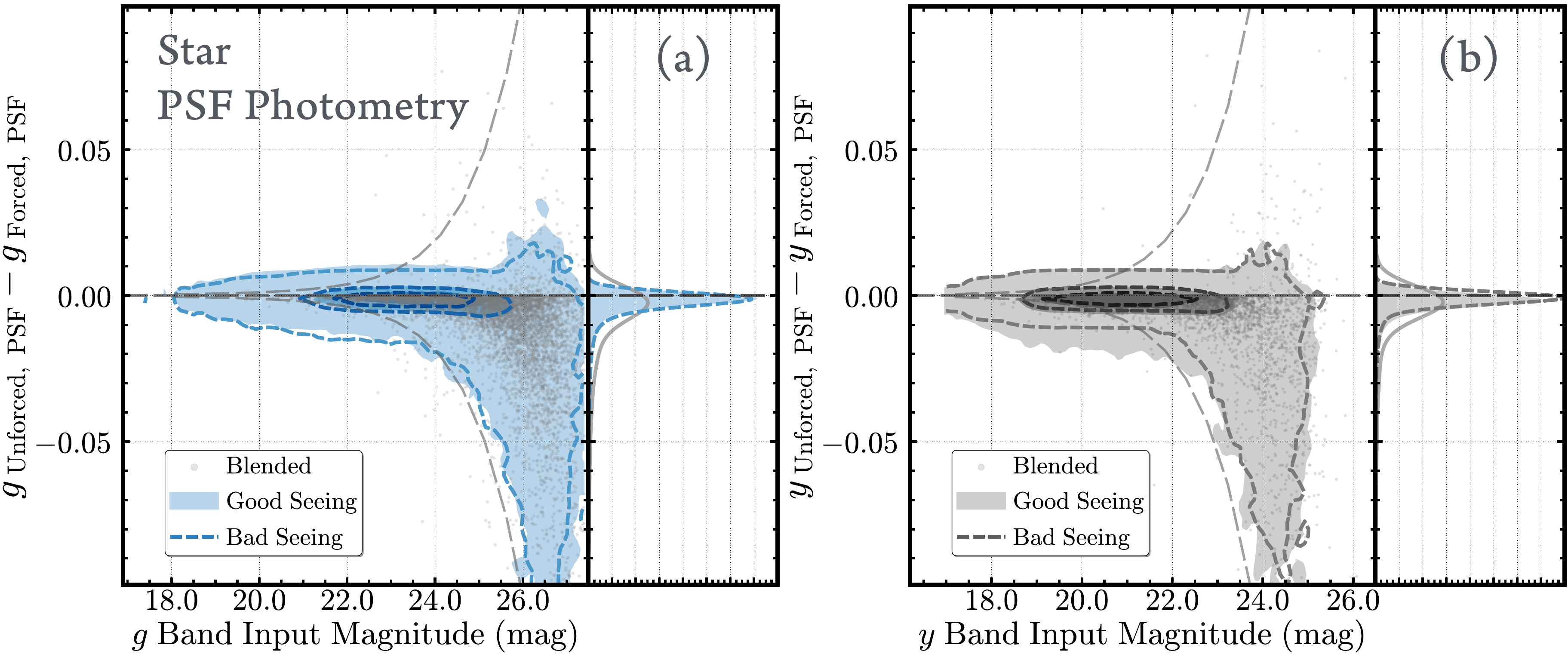}
    \end{center}
    \caption{
        The magnitude differences between the \unforced{} and \forced{}
        PSF photometry for synthetic stars in $g$ (\textbf{left}) and $y$ band
        (\textbf{right}).
        The lines and contours legend is identical to the plots in Fig \ref{fig:psf_mag}.
        }
    \label{fig:psf_diff}
\end{figure*}

\subsection{PSF photometry of stars}
    \label{ssec:psf}

    In \hscpipe{}, the PSF magnitude is derived using a matched-filter method that
    depends on the best-fit PSF model and uses third-order Lanczos interpolation to 
    shift the PSF model.
    The error of the PSF magnitude from \hscpipe{} only considers the per-pixel noise,
    and does not include centroid uncertainties.
    \hscpipe{} also estimates aperture correction for PSF photometry.
    For more details please see Bosch\etal (in prep.).

    We randomly inject ${\sim}100,000$ stars into each \tract{} in all five bands. 
    Typically, there are ${\sim}80000$ stars with $i_{\mathrm{PSF}}<26.0$ in one HSC 
    \tract{}. 
    Approximately ${\sim}3$--$4$\% of the stars are located within 2 pixels of the 
    centroids of real objects; we remove those stars from the sample to avoid
    confusion in the comparisons.
    For matched stars, we select the primary detections 
    (\texttt{detect.is-primary=True}) that have good photometric quality
    (see Appendix \ref{app:qc} for details).
    This gives us ${\sim}83,000$ stars in each \tract{}.

    Next, we exclude stars that are misidentified by \hscpipe{} as extended
    objects using the \texttt{classification.extendedness} parameter in each band.
    The fraction of misclassification is ${\sim}10$--$20$\%, and clearly depends on
    seeing conditions.
    For the same reason, the $g$-band has the highest fraction of misclassification,
    while $i$-band is recommended for selecting point sources.
    The star--galaxy separation issue is discussed more in Section \ref{ssec:sg}.

    In the following comparisons, we show the results from the
    \texttt{goodSeeing} and \texttt{badSeeing} \tracts{} separately because it
    is important to understand how the seeing conditions affects photometric accuracy 
    as well as to test whether \hscpipe{} can deliver unbiased photometry under 
    different seeing conditions.
    
    Besides seeing, the degree to which an object is blended with other objects 
    is another factor that influences photometric accuracy. 
    To quantify the impact of blending effects, we use the ``blendedness'' 
    parameter, $b$. 
    This parameter describes how much any given object overlaps with neighboring 
    objects (see Appendix \ref{app:defineb} and Murata\etal in prep.). 
    We divide our sample according to the blendedness parameter. 
    A value of $b=0$ corresponds to isolated objects whereas objects with $b=1$ 
    complete overlap with other objects. 
    Here we use $b>0.05$ to define highly blended stars (${\sim}4$--$9$\% are blended
    according to this criterion).
    The fraction of highly blended stars slightly increases in the \texttt{badSeeing} 
    \tract{}. 
    The impact of blendedness will be discussed more in Section \ref{ssec:blendedness}.

\subsubsection{Relationship between stellar magnitude and \s2n{}}

    The \s2n{} for PSF photometry is defined as
    $\mathrm{Flux}_{\mathrm{PSF}}/\mathrm{Flux\_Err}_{\mathrm{PSF}}$ in each band. 
    Figure \ref{fig:star_sn} shows the relationship between stellar magnitude and \s2n{} 
    in the $g$- and $i$-bands. 
    This figure shows the expected \s2n{} of stars as a function of magnitude in the 
    Wide layer.
    The slopes and scatters of these relations are very similar to the ones using 
    real stars on these two \tracts{}.
    
    Fig \ref{fig:star_sn} show that seeing conditions impact the \s2n{} of point
    sources at fixed input magnitude.
    In $i$-band, the \texttt{badSeeing} (0.70\asec{}) \tract{} shows systematically
    lower \s2n{} than the \texttt{goodSeeing} (0.45\asec{}) \tract{}.
    The \s2n{} is similar for the two $g$-band \tracts{} because they share similar 
    seeing conditions. 

    At \s2n{}$=5$, HSC Wide can detect stars as faint as ${\sim}26.5$ mag in both $g$ 
    and $i$ bands, which is consistent with the values found in \citet{HSCDR1} for 
    average seeing conditions.
    However, it is worth reminding HSC data users that the detection
    limit will exhibit spatial variations due to the seeing conditions.

\begin{figure*}
    \begin{center}
        \includegraphics[width=15cm]{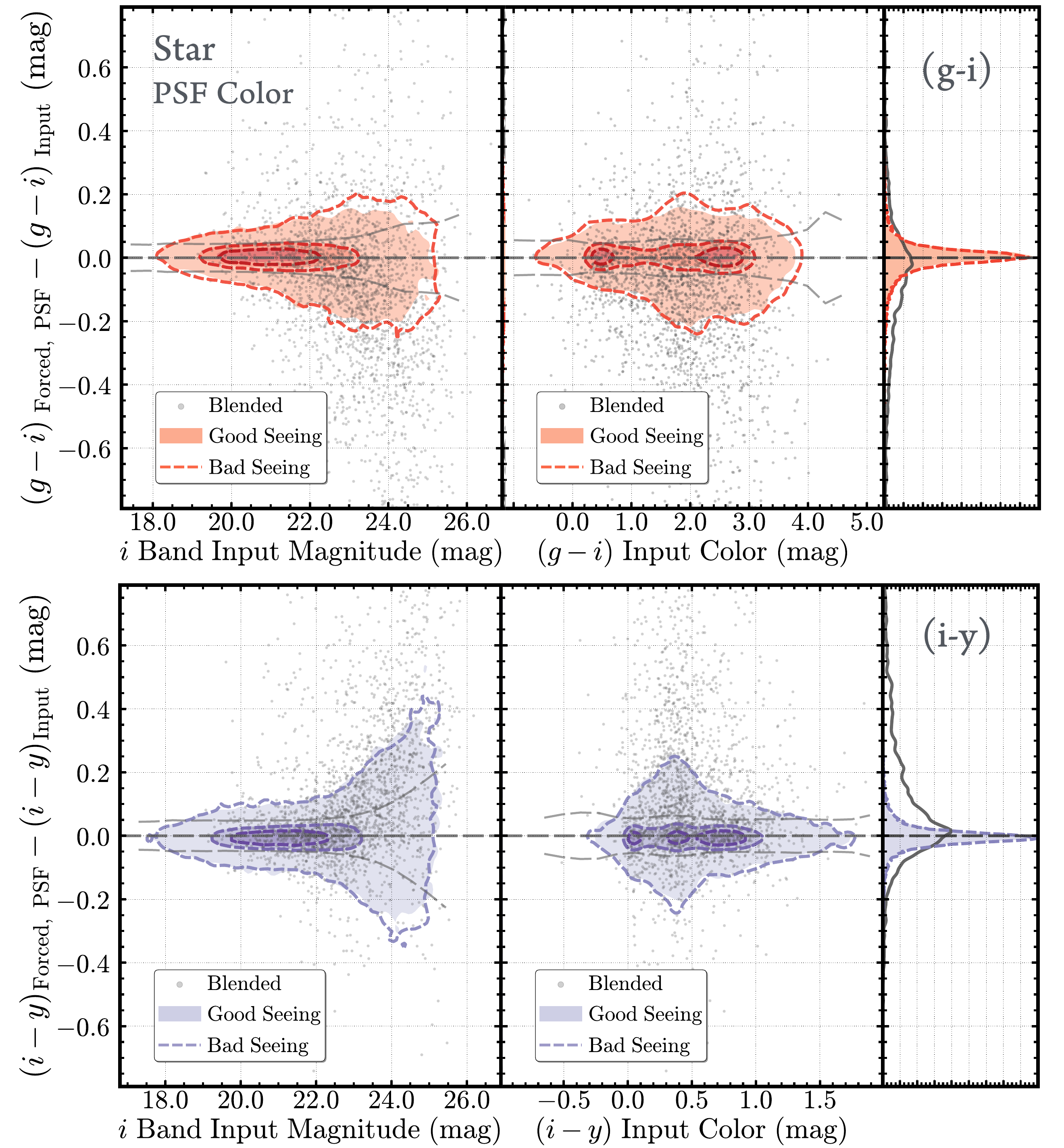}
    \end{center}
    \caption{
        Accuracy of the color measurements for synthetic stars via the differences
        between input and \forced{} PSF colors.
        The \textbf{upper panels} and \textbf{lower panels} are for $g-i$ and $i-y$
        colors separately.
        The \textbf{left} column shows the relation between input magnitude and
        the color difference, and the \textbf{right} column uses the input colors as
        x--axis instead.
        The lines and contours legend is identical to the plot in Fig \ref{fig:psf_mag}.
        }
    \label{fig:psf_color}
\end{figure*}

\begin{figure*}
    \begin{center}
        \includegraphics[width=\textwidth]{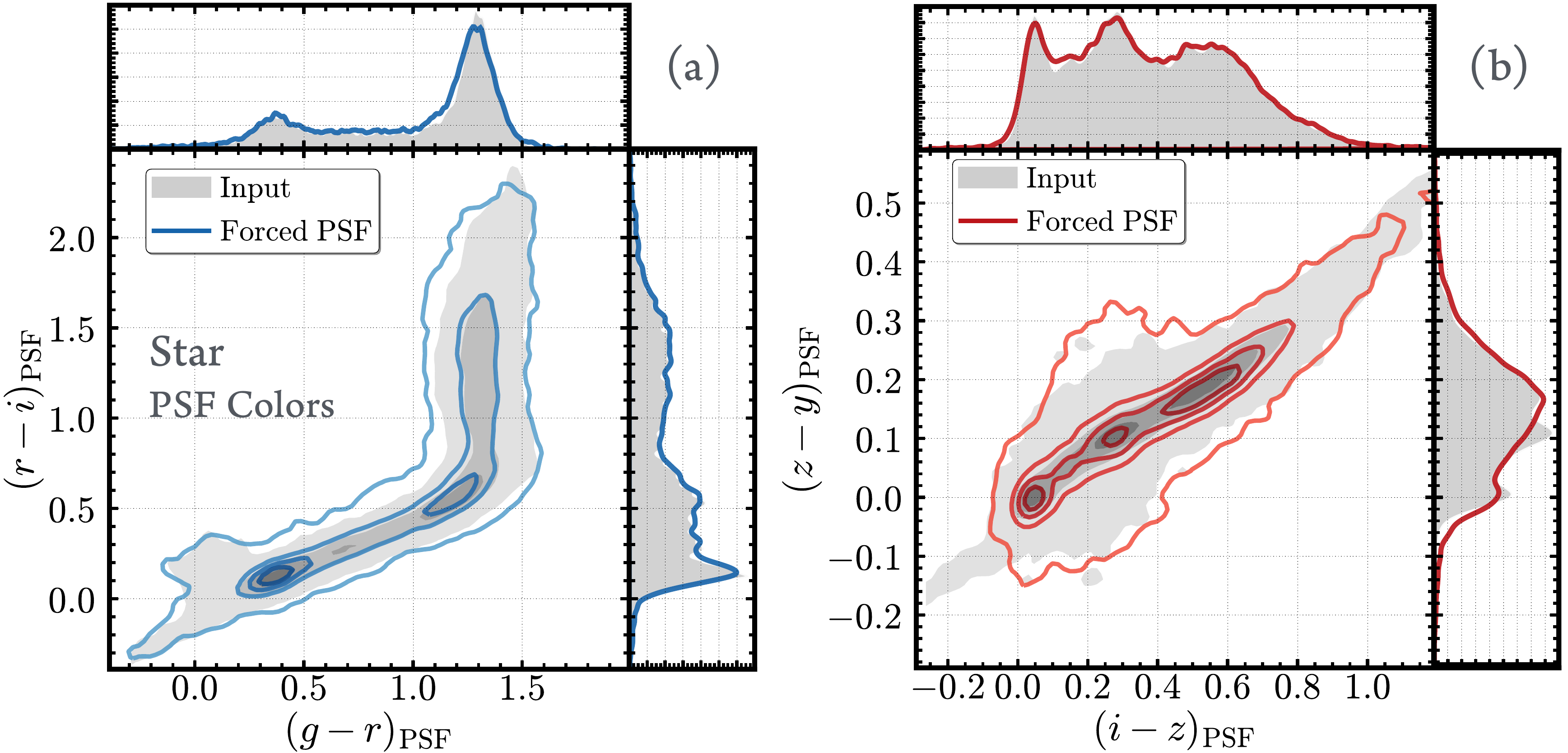}
    \end{center}
    \caption{
        Evaluations of color measurement accuracy for synthetic stars using
        the color-color distributions.
        The plot on the \textbf{left} (a) uses $(g-r)$ v.s. $(r-i)$ colors, and the plot 
        on the \textbf{right} (b) uses $(i-z)$ and $(z-y)$ colors.
        The filled contours and shaded histograms reflect the distributions for input
        colors.
        The empty contours and solid-line histograms show the distributions recovered
        by \hscpipe{}.
        }
    \label{fig:psf_cdist}
\end{figure*}

\begin{figure*}
    \begin{center}
        \includegraphics[width=\textwidth]{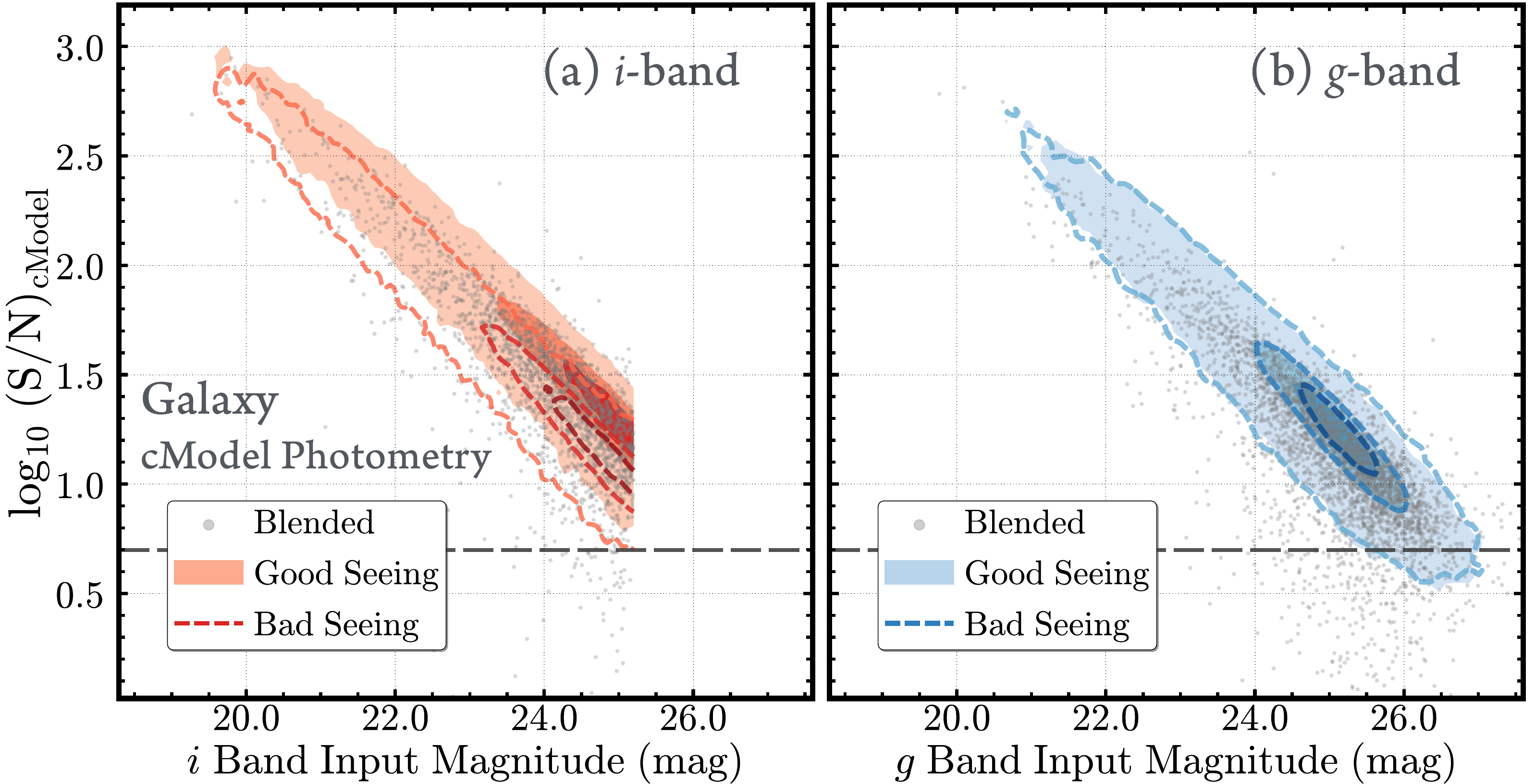}
    \end{center}
    \caption{
        Relation between input magnitudes (\textbf{left}: $i$-band; \textbf{right}:
        $g$-band) of synthetic galaxies $\log (\mathrm{S}/\mathrm{N})$ as measured 
        by \hscpipe{} \cmodel{} photometry. 
        Lines and contours are similar to Fig \ref{fig:star_sn}.
        The truncation in $i$ band input magnitude is caused by the magnitude limit 
        of COSMOS galaxies used in this test. 
        }
    \label{fig:cmodel_sn}
\end{figure*}

\subsubsection{Precision and Accuracy of PSF magnitudes}

    We now investigate the performance of the PSF photometry in each of the five bands
    independently. 
    Fig \ref{fig:psf_mag} shows the difference between the input magnitude versus the
    \hscpipe{} \forced{} PSF magnitude ($\Delta \mathrm{mag}_{\mathrm{PSF}}$) as a 
    function of input magnitude. 
    We separate these stars into seven input magnitude bins.  
    In each bin, we characterize the statistical \textbf{precision} of the PSF 
    magnitude using the standard deviation of the distribution (\smag{}).
    Meanwhile, we characterize the statistical \textbf{accuracy} of the PSF magnitude 
    via the mean magnitude difference in each bin (\mmag{}). 
    This also informs us whether the PSF photometry is biased at certain input 
    magnitude.
    
    The overall performance of the \hscpipe{} \forced{} PSF magnitude is excellent. 
    The median $i$-band PSf magnitude precision for the \texttt{goodSeeing} \tract{} 
    is ${\sim}0.014$ mag (${\sim}1.3$\%) at $i_{\mathrm{Input}}{\sim}19.0$ mag. 
    At $i_{\mathrm{Input}}{\sim}24.0$ mag, the precision of the PSF magnitude is 
    at ${\sim}3$\% level (${\sim}0.030$ mag statistical scatter).
    At $i_{\mathrm{Input}}{\sim}25.0$ mag, the precision decreases to
    ${\sim}6$\% with an average ${\sim}0.062$ mag difference.
    
    The precision of PSF magnitude for the \texttt{badSeeing} \tract{} shows similar 
    performance at $i_{\mathrm{Input}}<23.5$ mag. 
    At fainter magnitudes, the precision slightly degrades. 
    The \texttt{badSeeing} \tract{} has ${\sim}4$\% precision at
    $i_{\mathrm{Input}}{\sim}24.0$ mag and ${\sim}11$\% precision at
    $i_{\mathrm{Input}}{\sim}25.0$ mag.
    \citet{HSCDR1} evaluates the precision of PSF magnitude via external comparisons
    with the PS1, PV2, and SDSS data at $i<21$ mag, and finds it at 1-2\% level,
    which is consistent with our results. 
    However, at fainter magnitudes, external comparisons become difficult due to the 
    lack of imaging matched to HSC depths.
    The results from \synpipe{} hence provide useful evaluations of the precision of 
    the HSC photometry down to our detection limit\footnote{Strictly speaking, the 
    precision reported here should be considered as upper limits because 
    \synpipe{} currently does not consider the systematic uncertainties in PSF 
    modeling. But this effect should be subdominant.}

    Fig \ref{fig:psf_mag} also shows that the precision of the PSF photometry is 
    not filter dependent. 
    The precision of \forced{} PSF magnitudes in $r$ and $z$ is similar to the results
    for $i$-band. 
    For $r$ band we find 1.5-4.0\% precision at $r_{\mathrm{Input}} < 24.0$ mag and 
    ${\sim}8$\% precision at $r_{\mathrm{Input}}{\sim}25.0$. 
    For $z$-band we find 1.0-5.0\% precision at $z_{\mathrm{Input}} < 24.0$ mag and 
    ${\sim}10$\% precision down to $z_{\mathrm{Input}}{\sim}25.0$ mag).
    
    The precision for $g$ and $y$ bands becomes slightly worse at the very faint end. 
    For $g$ band, the precision is 1.5-7.0\% at $g_{\mathrm{Input}} < 24.0$ mag and is
    ${\sim}13$\% down to $25.0$ mag.
    For $y$ band, we find 1.5-5.0\% precision at $y_{\mathrm{Input}} < 23.0$ mag and
    ${\sim}12$\% down to $24.0$ mag. 
    These differences are however consistent with the differences in the seeing 
    conditions between the filters 
    

    The PSF photometry for relatively isolated stars is accurate and unbiased in $i$ 
    and $z$ bands down to faint magnitudes. 
    For the $g$, $r$, and $y$ bands, we find mean $\Delta \mathrm{mag}_{\mathrm{PSF}}$
    that are close to the PSF flux errors estimated by \hscpipe{}.  
    \hscpipe{} tends to underestimate the fluxes of stars in these bands by 
    ${\sim}0.01-0.02$ mag at $< 24.0$ mag. 
    The exact cause of this offset is unclear but the levels of bias is quite small 
    and not a major concern.  

	However, Fig \ref{fig:psf_mag} also shows that blending has a strong impact on 
	photometry.  
	On average, \hscpipe{} systematically underestimates the total fluxes of stars 
	that are subject to blending effects by $0.05$--$0.10$ mag at a fixed input 
	magnitude.
    We see the same effect in all bands, and the impact of blendedness becomes
    increasingly significant at fainter magnitudes.
    It is important to bear this caveat in mind when using PSF photometry for point
    sources in HSC data.

	Finally, we also perform similar tests for the \unforced{} PSF photometry in all 
	five bands, and find similar results. 
	To measure the \forced{} PSF photometry, \hscpipe{} fixes the centroid of the PSF 
	model across all five bands. 
	A difference between \forced{} and \unforced{} PSF magnitudes would therefore be 
	an indication of photometric uncertainties arising from inaccurate astrometric 
	calibrations and PSF modeling across different bands\footnote{On average, the 
	centroids are more accurate in \forced{} photometry as they are defined using 
	the band with higher \s2n{}}. 
	Fig \ref{fig:psf_diff} shows the difference between \forced{} and \unforced{} 
	PSF photometry for the $g$ and the $y$ band. 
	The overall differences are small ($<0.01$ mag). 
	At very faint end ($g_{\mathrm{Input}}>24.0$ mag and $y_{\mathrm{Input}}>23.5$,
	the \unforced{} photometry slightly overestimate the fluxes of stars comparing 
	to the \forced{} photometry. 
	
	The precisions and accuracies of \forced{} PSF magnitudes shown here are 
	summarized in Table \ref{tab:psfmag}

\subsubsection{Precision and Accuracy of PSF colors}

    PSF photometry is the most appropriate way to measure colors for point
    sources; therefore, the accuracy of the PSF color estimates is important to many 
    scientific goals of the HSC survey
    (e.g., study of the Milky Way structure, selection of unique stellar objects or
    high-redshift quasars, and accurate star--galaxy separation).

    In Fig \ref{fig:psf_color}, we evaluate the precision and accuracy of the 
    \forced{} $(g-i)$ and $(i-y)$ PSF colors by comparing the differences between input 
    and output colors ($\Delta \mathrm{Color}$) with both the input magnitudes and 
    colors. 
    Fig \ref{fig:psf_color} shows that \hscpipe{} provides precise and accurate PSF 
    colors for synthetic stars with realistic color distributions down to very 
    faint magnitudes. 
    The average precision for $(g-i)$ is ${\sim}0.023$ mag at 
    $i_{\mathrm{Input}}{\sim}19.0$. 
    The statistical scatter increases to ${\sim}0.16$ mag at 
    $i_{\mathrm{Input}}{\sim}25.0$ mag.  
    The $(i-y)$ color displays similar precision. 
    The \forced{} PSF color measurements are not biased through the entire ranges of 
    input magnitudes and colors.
    
    Fig \ref{fig:psf_color} shows that the precision of PSF colors do not depend 
    the seeing conditions. 
    For instance, the precision of $(g-i)$ for the \texttt{badSeeing} \tract{} is 
    only slightly worse at the very faint end (${\sim}0.18$ mag at
    $i_{\mathrm{Input}}{\sim}25.0$ mag) compared to the \texttt{goodSeeing} \tract{}.

    Fig \ref{fig:psf_cdist} displays the precision of PSF colors by comparing the 
    color distributions of the input sample to the recovered color distributions. 
    Using four different colors and two color--color planes, we show that the 
    \forced{} PSF colors accurately recover the distributions of all four colors 
    and the color-color distributions.

    As for highly blended stars, Fig \ref{fig:psf_color} also shows that highly blended 
    stars have worse precision and accuracy in their colors compared to isolated stars. 
    At fixed input magnitude, the precision in $(g-i)$ is ${\sim}0.1$ mag worse for 
    highly blended stars.  
    Also, for blended stars, $(g-i)$ shows a bias towards bluer colors, while $(i-y)$ 
    shows a bias towards redder colors. 
    However, such biases appear to be less severe in terms of colors than in terms of
    PSF magnitudes.
    
	The statistical uncertainties of \forced{} PSF colors are summarized in 
	Table \ref{tab:psfcolor}
    

\begin{figure*}
    \begin{center}
        \includegraphics[width=16cm]{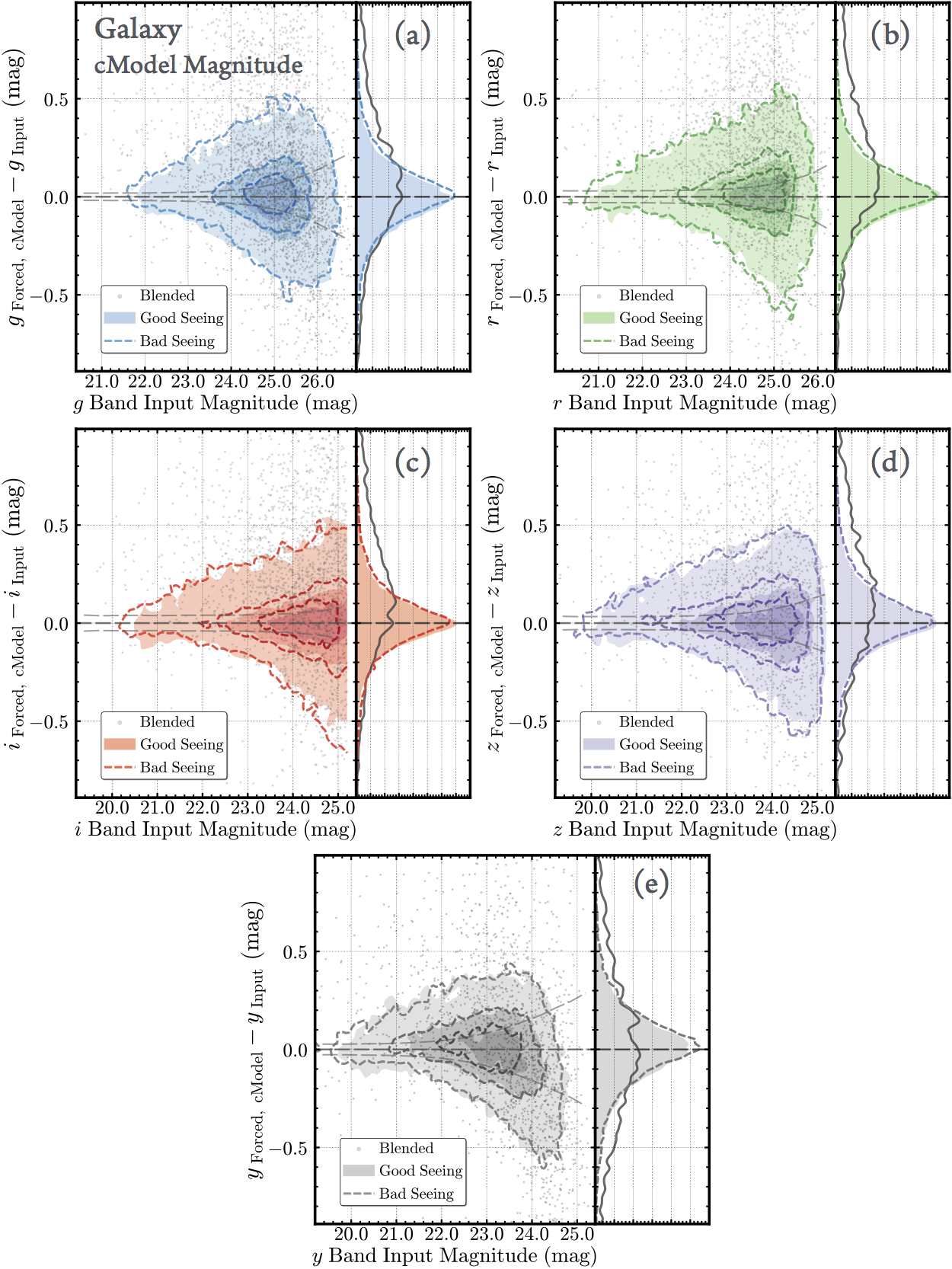}
    \end{center}
    \caption{
        Accuracies of the \hscpipe{} \cmodel{} photometry for synthetic
        galaxies measured by the difference between input and output \forced{}
        \cmodel{} magnitudes.
        Plots [a, b, c, d, e] show the results for [$g$, $r$, $i$, $z$, $y$]-bands, 
        respectively.
        The lines and contours legend is identical to Fig \ref{fig:psf_mag}.
        }
    \label{fig:cmodel_mag}
\end{figure*}

\begin{figure*}
    \begin{center}
        \includegraphics[width=\textwidth]{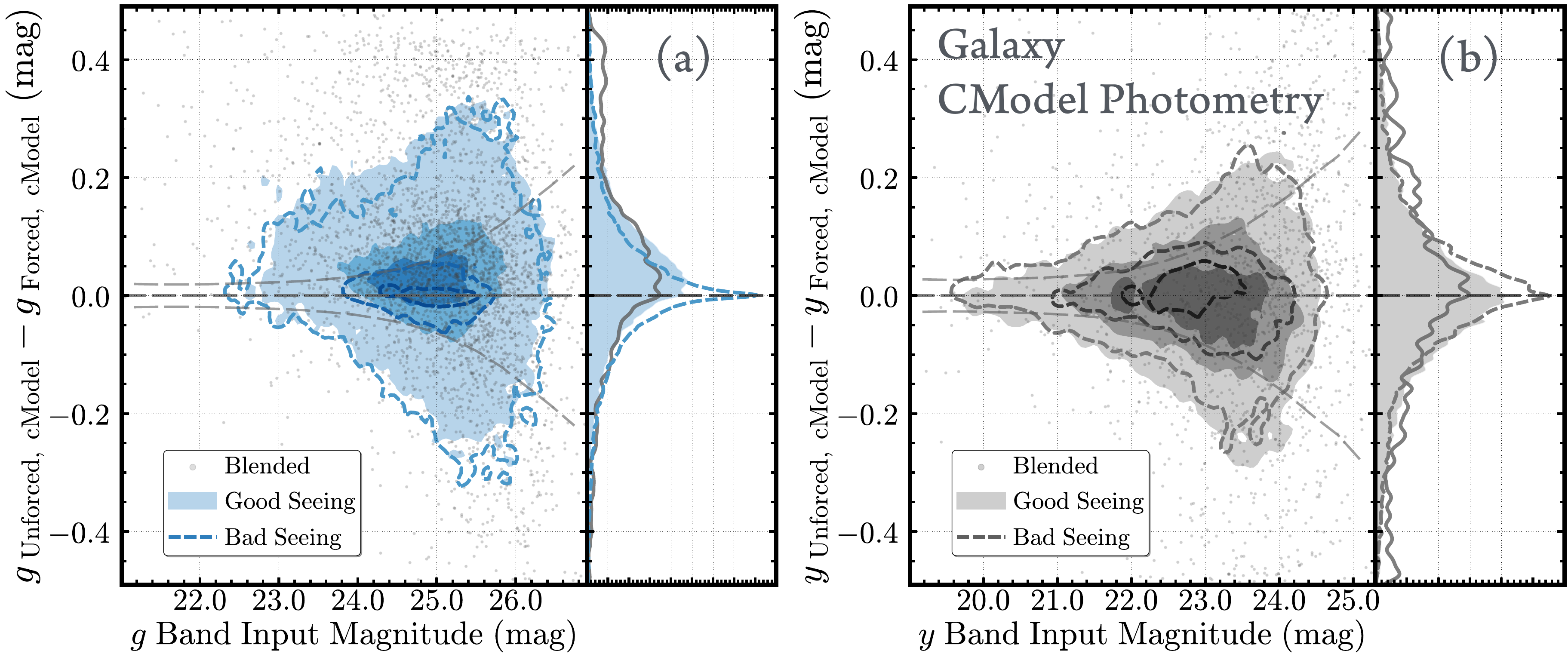}
    \end{center}
    \caption{
        The magnitude differences between the \unforced{} and \forced{}
        \cmodel{} photometry for synthetic galaxies in $g$ (\textbf{left}) and
        $y$ band (\textbf{right}).
        The lines and contours legend is identical to the plots in Fig 
        \ref{fig:psf_diff}.
        }
    \label{fig:cmodel_diff}
\end{figure*}

\begin{figure*}
    \begin{center}
        \includegraphics[width=15cm]{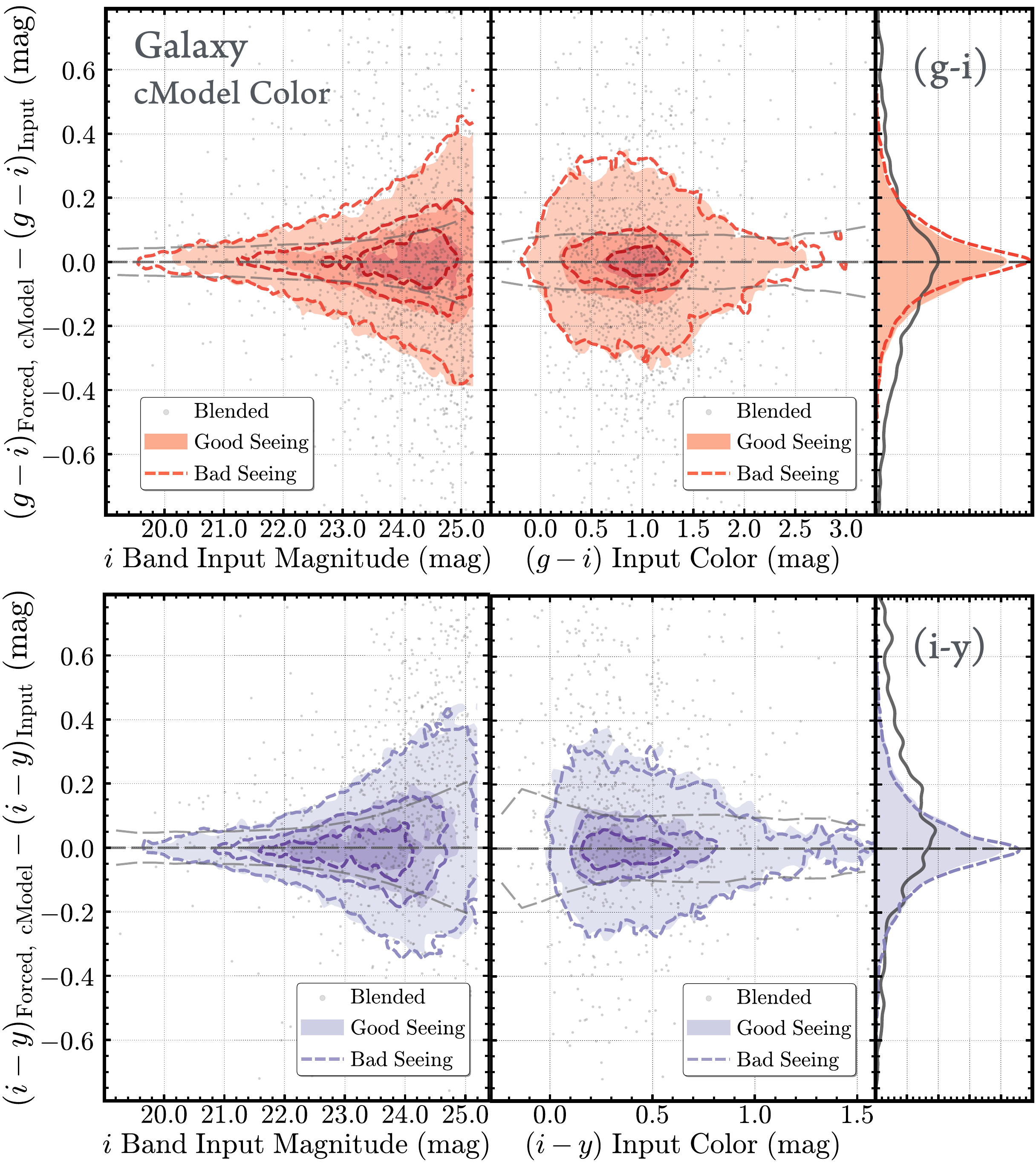}
    \end{center}
    \caption{
        Accuracies of the color measurements for synthetic galaxies via the
        differences between input and \forced{} \cmodel{} colors.
        The \textbf{upper panels} and \textbf{lower panels} are for $g-i$ and $i-y$
        colors separately.
        The \textbf{left} column shows the relation between input magnitude (x--axis) and
        the color difference, and the \textbf{right} column uses the input colors as
        the x--axis.
        The lines and contours legend is identical to the plot in 
        Fig \ref{fig:psf_color}.
        }
    \label{fig:cmodel_color}
\end{figure*}

\begin{figure*}
    \begin{center}
        \includegraphics[width=\textwidth]{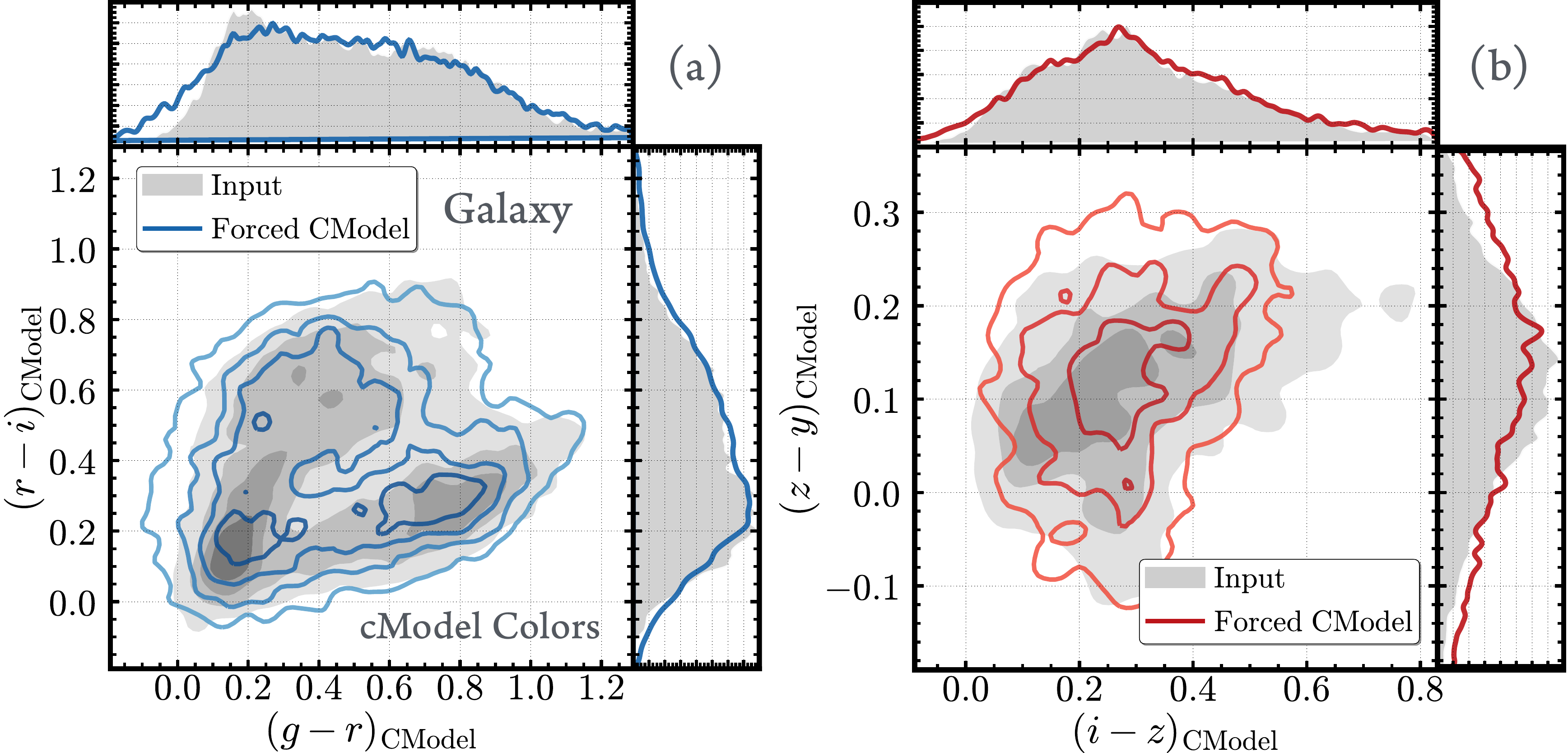}
    \end{center}
    \caption{Color-color distributions of synthetic  galaxies.
        The \textbf{left} panel (a) uses $(g-r)$ vs. $(r-i)$ colors, and the 
        \textbf{right} panel (b) uses $(i-z)$ and $(z-y)$ colors. 
        Filled contours and shaded histograms correspond to the input colors. 
        Empty contours and solid-line histograms show the distributions recovered
        by \hscpipe{} \cmodel{} photometry.
        }
    \label{fig:cmodel_cdist}
\end{figure*}

\subsection{cModel photometry of Galaxies}
    \label{ssec:cmodel}

    For galaxies, \hscpipe{} uses a \cmodel{} photometry algorithm that is an improved 
    version of the SDSS \cmodel{} photometry.
    For details about the \cmodel{} algorithm, please see description in Section 
    \ref{ssec:hscpipe} and Bosch\etal (in prep.).
    Despite the limitations of the \cmodel{} method (e.g., sensitivity to the 
    background subtraction and to deblending failures), it can deliver robust 
    PSF-corrected fluxes and colors for galaxies.

    Typical \tract{} in the Wide layer contains ${\sim}400000$ extended objects with 
    $i_{\mathrm{CModel}} < 25.5$ mag. 
    We randomly inject ${\sim}40,000$ synthetic galaxies (additional ${\sim}10$\%) 
    into each \tract{}, and will not create over-crowded situation.
    Using similar approach for synthetic stars, we select galaxy samples for 
    photometric comparisons. 
    Synthetic galaxies that are located within 2 pixels of the centroids of real 
    objects are removed from the sample (they represent ${\sim}3$--$7$\% of the 
    input sample). 
    After imposing quality cuts on the recovered synthetic galaxy sample 
    (see Appendix \ref{app:qc} for details), we have ${\sim}30,000$ galaxies in 
    each \tract{}.
    
    We find that \hscpipe{} mis-classifies lower than $1$\% of these objects as 
    point sources in the \texttt{goodSeeing} \tract{}.
    But the fraction of misclassified galaxies increases to $4$\% in the
    \texttt{goodSeeing} \tract{}. 
    We will discuss star/galaxy separation further in Section \ref{ssec:sg}.

    We define a sample of highly blended galaxies by imposing the cut 
    $b>0.05$\footnote{We already removed the ambiguously blended objects. 
    These objects have high blendedness, but are still distinctive.}.
    Due to the extended nature of galaxies, the $b$ distribution of synthetic 
    galaxies is more skewed toward high values (especially for the \texttt{badSeeing}
    \tract{}) compared to the b distribution of stars.
    About $7$--$9$\% of synthetic galaxies are highly blended with $b>0.05$.
    The fraction is slightly higher in the \texttt{badSeeing} \tract{}, and 
    therefore the $g$ band also has higher fraction of highly blended galaxies.
    In our plots, we will highlight highly blended galaxies and will discuss 
    blending effects further in Section \ref{ssec:blendedness}.

\subsubsection{Input magnitude and the \s2n{} of \cmodel{} photometry}

    Fig \ref{fig:cmodel_sn} shows the relation between the input magnitudes and the 
    \s2n{} of synthetic galaxies.  
    The \s2n{} is the ratio  of the \cmodel{} flux and the flux error measured by 
    \hscpipe{}.
    Here the adopted \s2n{} corresponds to the average \s2n{} over entire footprint. 
    The center of the galaxy typically has high \s2n{} than this value.    

    Fig \ref{fig:cmodel_sn} shows that the \cmodel{} photometry displays a 
    well-behaved relation between input magnitude and output \s2n{}.  
    Compared to stars, the \s2n{} from \cmodel{} for synthetic galaxies shows a larger
    scatter at fixed input magnitude. 
    This is likely due to the fact that galaxies have a wider range in sizes and shapes 
    than point sources.  
    In $i$ band, a typical $25.2$ mag galaxy has \s2n{}${\sim}20$ from \cmodel{}, but 
    there are also galaxies at the same magnitude which have \s2n{} below 5. 
    The HSC Wide layer can detect galaxies as faint as $i{\sim}26.0$ mag but the 
    sample becomes incomplete at these faint magnitudes.
    This also introduces a selection bias if galaxies with certain structural 
    properties (e.g., more extended and lower surface brightness) are harder to detect
    than others.
    HSC survey data users who want to select flux-limited galaxy samples,
    or who intend to study the population of faint (e.g., high-$z$) galaxies should
    keep this in mind.  
    Finally, we also find that a worse seeing leads to a lower \s2n{} at fixed 
    magnitude (see Fig \ref{fig:star_sn} for $i$-band).

\subsubsection{Precision and Accuracy of the \cmodel{} magnitude}

    Fig \ref{fig:cmodel_mag} shows the precision (\smag{}) and accuracy (\mmag{}) of 
    \cmodel{} magnitude in each band using the same format as Fig \ref{fig:psf_mag}.

    The overall performance of \cmodel{} photometry is reasonable down to 
    $i_{\mathrm{Input}}=25.2$ mag.
    Compared to the PSF photometry for stars, the statistical uncertainties of \cmodel{} 
    magnitudes are larger for galaxies, which is expected given the diversity of galaxy 
    shapes and sizes that adds complexity to the model-fitting process.
    At the same time, Fig \ref{fig:cmodel_mag} shows that the \cmodel{} algorithm in
    \hscpipe{} provides unbiased and consistent photometry for galaxies across 
    different bands and seeing conditions.

    The typical precision of $i$-band flux using \forced{} \cmodel{} is at 
    ${\sim}10-14$\% level in the \texttt{goodSeeing} \tract{} at 
    $i_{\mathrm{Input}}<24.0$ mag. 
    It moderately degrades to ${\sim}18$\% at $25.0$ mag. 
    The performance is similar in $i$ band for the \texttt{badSeeing} \tract{}. 
    At $20.0 < i_{\mathrm{Input}} < 24.0$ mag, the accuracies of \forced{} \cmodel{}
    magnitudes change between \plus{}$0.017$ mag to \minus{}$0.023$ mag. 
    At $24.0 < i_{\mathrm{Input}} < 25.5$ mag, \mmag{} is around $\minus0.006$ mag,
    which suggest that \forced{} \cmodel{} photometry is unbiased down to the very 
    faint end.
    
    It is worth noting that the flux uncertainty from \synpipe{} are significantly 
    larger than the flux errors from \hscpipe{}, which does not take into account the 
    systematic uncertainties involved in the modeling fitting process. 
    Even though our synthetic galaxy sample is comprised of objects with simple 
    single-\ser{} models, \cmodel{} algorithm can only approximate them to a certain 
    accuracy because of the background noise, deblending uncertainties, and priors 
    imposed on model parameters. 
    Given that real galaxies are more complex in structure, our quoted uncertainties 
    should be treated as lower limits.


    The precision of \forced{} \cmodel{} photometry is consistent across all filters.
    For the $g$, $r$, and $z$ bands, the statistical uncertainties are very similar to
    $i$-band results between $20.0$ and $25.0$ mag.
    The \forced{} \cmodel{} magnitudes are also unbiased for these three filters
    across the entire input magnitude range.
    The precision for the $y$ band is slightly worse, ranging from 10\% to 17\% in the
    same magnitude bins.
    In addition, the \forced{} \cmodel{} tends to overestimate the total flux in 
    the $y$ band at $>24.0$ mag. 
    This bias in the $y$ band flux is equal to \minus{}0.07 at $24.0$ mag to 
    \minus{}0.22 mag at 25.0 mag. 
    Worse seeing conditions and a higher background noise level mean that it is more
    difficult to detect faint objects in the $y$ band.
    At $y_{\mathrm{Input}}{\sim}24.0$, the average \s2n{} for \cmodel{} is already
    around the detection threshold.

    Our synthetic galaxy sample is dominated by faint galaxies 
    ($i_{\mathrm{Input}} > 24.0$ mag) that are crucial to key scientific goals of the 
    HSC survey.
    But it also means poor statistics for bright galaxies. 
    We do find that at $i_{\mathrm{Input}}<20.0$, the \forced{} \cmodel{} photometry 
    starts to systematically underestimate total fluxes. 
    Based on external comparisons, the behavior of \cmodel{} for bright galaxies may 
    depend on galaxy types (e.g., \citealt{HSCDR1}).  
    Also, at bright end, another known issue is that the current \hscpipe{}
    over-deblends around bright objects. 
    Inappropriately weight priors in \cmodel{} also leads to poorer photometric 
    quality for bright galaxies.  
    Finally, at the imaging depth of HSC ($>29.0$\sb{} in the Wide Layer $i$ band),  
    \cmodel{} may simply not be a good choice for modeling bright galaxies 
    (e.g., massive elliptical galaxies; see Huang\etal in prep.~for more details).

    We also find that highly blended galaxies tend to have systematically 
    underestimated total fluxes (with an average offset ${\sim}0.1$ mag) and 
    higher statistical uncertainties in all five bands. 
    However, the contrast between relatively isolated and highly blended galaxies
    in photometric performance is not as stark as for synthetic stars 
    (Fig \ref{fig:psf_mag}).

    We also test the \unforced{} \cmodel{} photometry in all five bands, and the
    results suggest similar precision.
    Fig \ref{fig:cmodel_diff} compares the difference between \forced{} and \unforced{} 
    \cmodel{} in $g$ and $y$ bands.  
    The main noticeable trend is that galaxies brighter than $24.0$ mag in the $g$ band 
    have systematically brighter \forced{} \cmodel{} magnitudes, which suggests that 
    the \forced{} \cmodel{} is more accurate in the $g$ band as the model parameters 
    are determined using band with higher \s2n{}.
    
	The precisions and accuracies of \forced{} \cmodel{} magnitudes shown here are 
	summarized in Table \ref{tab:cmodelmag}

\subsubsection{Precision and Accuracy of the \cmodel{} color}

    Precise and unbiased five-band \cmodel{} colors from \hscpipe{} are crucial for
    photometric redshift estimates and spectral energy distribution (SED) fitting 
    results. 
    They are also key, for example, to the selection of high-$z$ Lyman-break 
    galaxies (LBG).

    Fig \ref{fig:cmodel_color} evaluates the precision and accuracy of \forced{} 
    $(g-i)$ and $(i-y)$ \cmodel{} colors using the same format as 
    Fig \ref{fig:psf_color}. 
    We find that \hscpipe{} provides reliable \cmodel{} colors for synthetic 
    galaxies down to $i_{\mathrm{Input}}{\sim}25.0$ mag.
    For $(g-i)$ colors, the statistical uncertainty at $i_{\mathrm{Input}}{\sim}20.0$ 
    mag is ${\sim}0.07$ mag.
    The statistical uncertainty increases to ${\sim}0.18$ mag at 
    $i_{\mathrm{Input}}{\sim}25.0$ mag.
    The $(i-y)$ colors show similar results, except the precision at 
    $i_{\mathrm{Input}}>24.0$ mag becomes slightly worse.

    The performance of \forced{} \cmodel{} color does not strongly depend on seeing 
    and is unbiased for the entire range of magnitudes and colors that we have tested.
    The only noticeable feature is that the $(i-y)$ color at $i_{\mathrm{Input}}>24.0$ 
    is systematically redder than the real values by $\plus0.05$ mag.
    As mentioned above, this is not surprising given the shallower $y$ band 
    imaging depth. 
    Highly blended galaxies are less precise in their \forced{} \cmodel{} colors 
    compared to isolated galaxies.  
    Highly blended galaxies on average show slightly bluer $(g-i)$ colors and redder  
    $(i-y)$ colors compared to their input colors.

    Fig \ref{fig:cmodel_cdist} compare the input color--color distributions with the
    recovered ones. 
    The general 2--D distributions are well recovered but the precision is not as 
    good as for stars, especially in the $(i-z)$ vs. $(z-y)$ plane.
    This is expected given the narrow dynamical ranges of these colors in the redder 
    filters.
    The \forced{} \cmodel{} tends to slightly underestimate the $(g-r)$ and $(i-z)$
    colors at the very ``blue'' end, while overestimating the $(i-y)$ color at the very
    red end.

    We note that our synthetic galaxies do not have color gradients.  
    In reality, color gradients will complicate the situation and hence the statistical 
    uncertainties shown here should be considered as lower limits.

	The statistical uncertainties of \forced{} \cmodel{} colors are summarized in 
	Table \ref{tab:cmodelcolor}
    

\begin{figure*}
    \begin{center}
        \includegraphics[width=\textwidth]{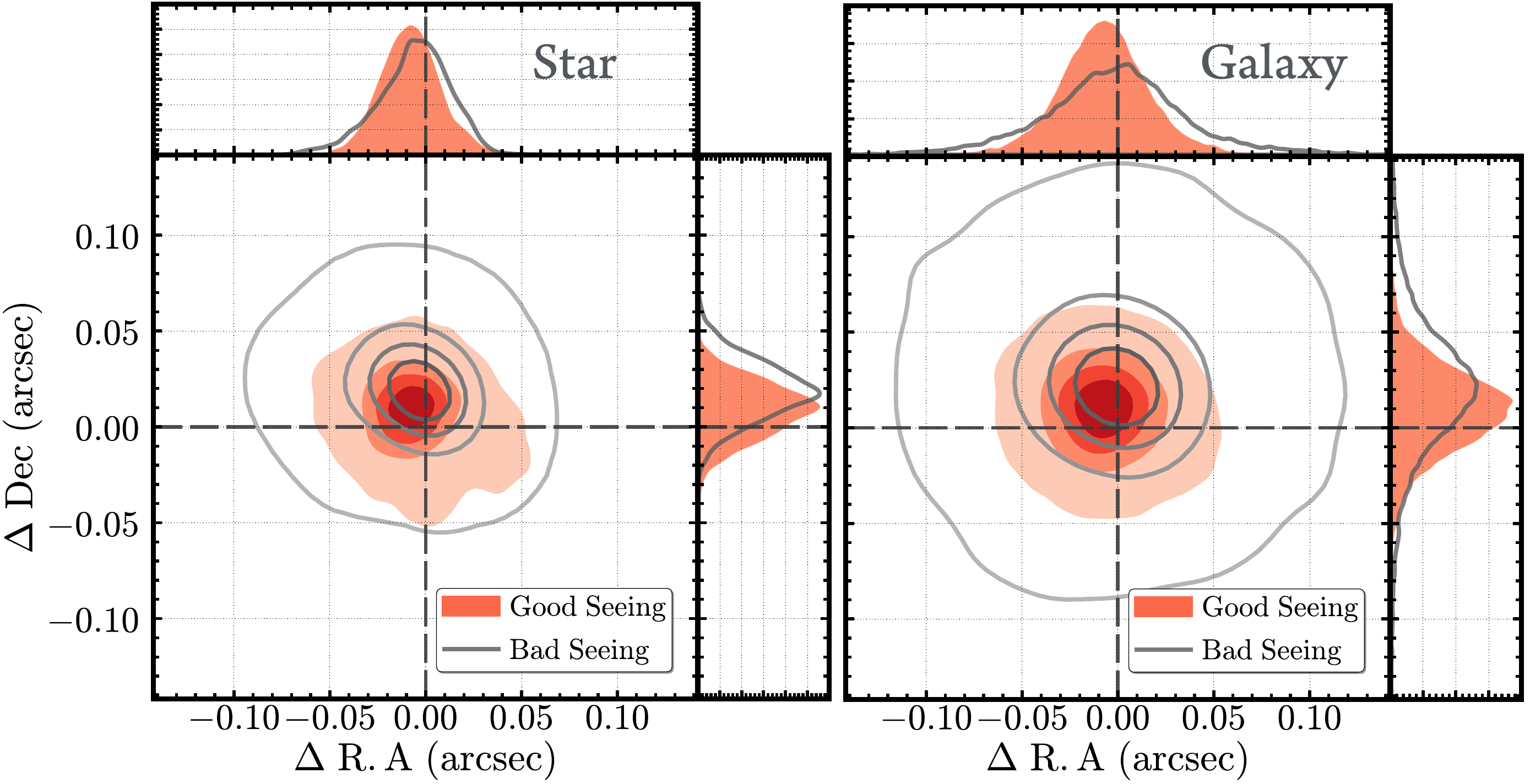}
    \end{center}
    \caption{
        Astrometric accuracy for synthetic stars (\textbf{left}) and galaxies
        (\textbf{right}). 
        We show the differences between the input R.A and Dec and the coordinates 
        measured on the \coadd{} images. 
        Filled-contours (filled histogram) and open-contours (solid-line histogram) are 
        for synthetic objects from \tracts{} with good and bad seeing conditions, 
        respectively.
        $\Delta\mathrm{R.A}=0$ and $\Delta\mathrm{Dec}=0$ are marked by dashed lines.     
        It shows that the initial astrometric calibration is already very accurate,  
        and it also suggests that \hscpipe{} can find the centroids of stars and 
        galaxies with precisions that are consistent with the astrometric calibration
        uncertainty under moderately different seeing conditions.
        }
    \label{fig:astrometry}
\end{figure*}

\begin{figure*}
    \begin{center}
        \includegraphics[width=\textwidth]{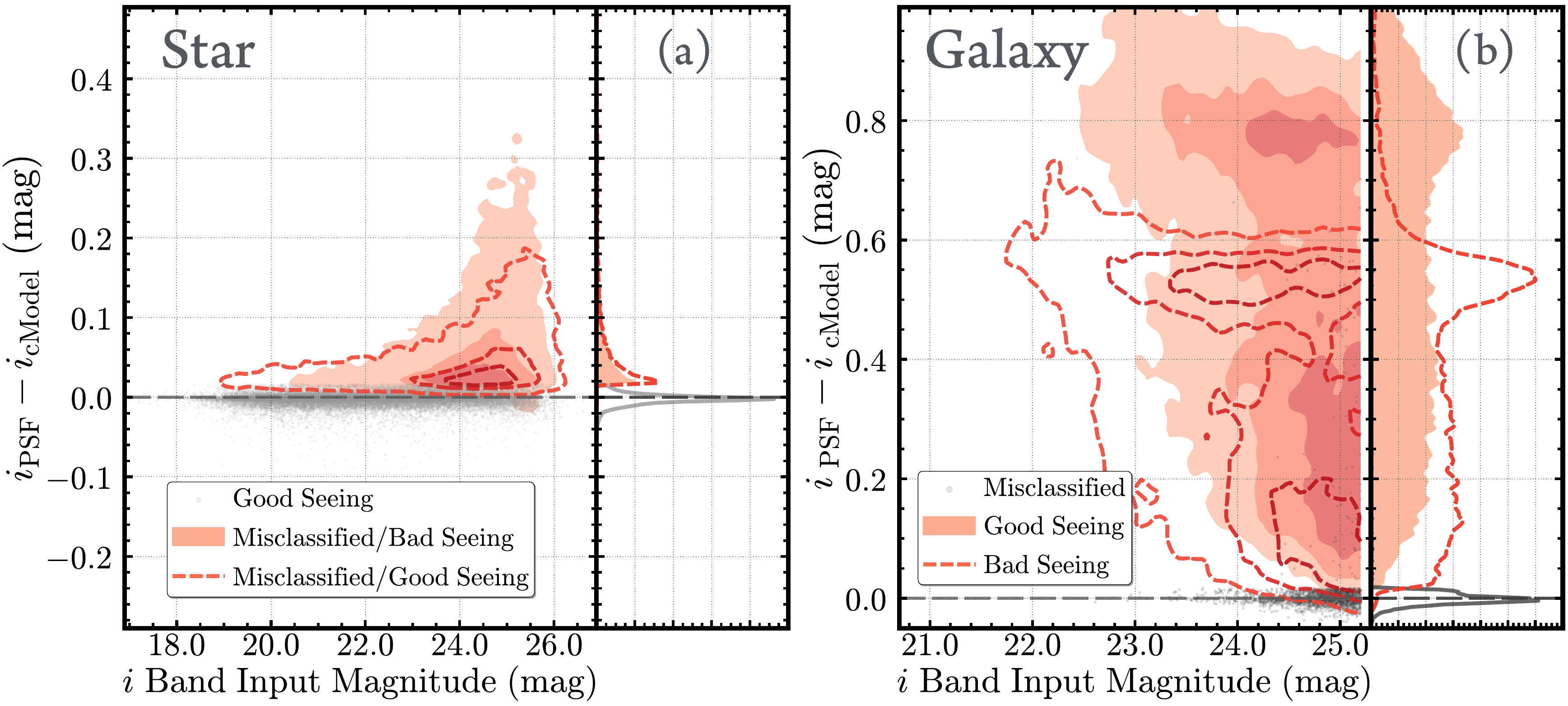}
    \end{center}
    \caption{Evaluation of the \hscpipe{} star--galaxy separation scheme. 
        Star--galaxy separation is based on the the magnitude difference between the 
        PSF and \cmodel{} photometry. 
        The plot on the \textbf{left} (a) corresponds to synthetic stars. 
        Filled-contours and filled histogram show distributions of correctly classified
        synthetic stars in the \texttt{goodSeeing} \tract{}.
        The scatter plot (dashed-line histogram) and open-contour (solid-line histogram)
        show distributions of synthetic stars that are misclassified as extended
        objects in \tracts{} with bad and good seeing conditions, respectively.
        The plot on the \textbf{right} (b) corresponds to synthetic galaxies.
        Filled-contours (filled histogram) and open-contours (solid-line histogram)
        are for galaxies from \texttt{goodSeeing} and \texttt{badSeeing} \tracts{},
        respectively.
        The scatter plot and dashed-line histogram highlight synthetic galaxies that 
        are misclassified as point sources.}
    \label{fig:sg}
\end{figure*}

\begin{figure*}
    \begin{center}
        \includegraphics[width=\textwidth]{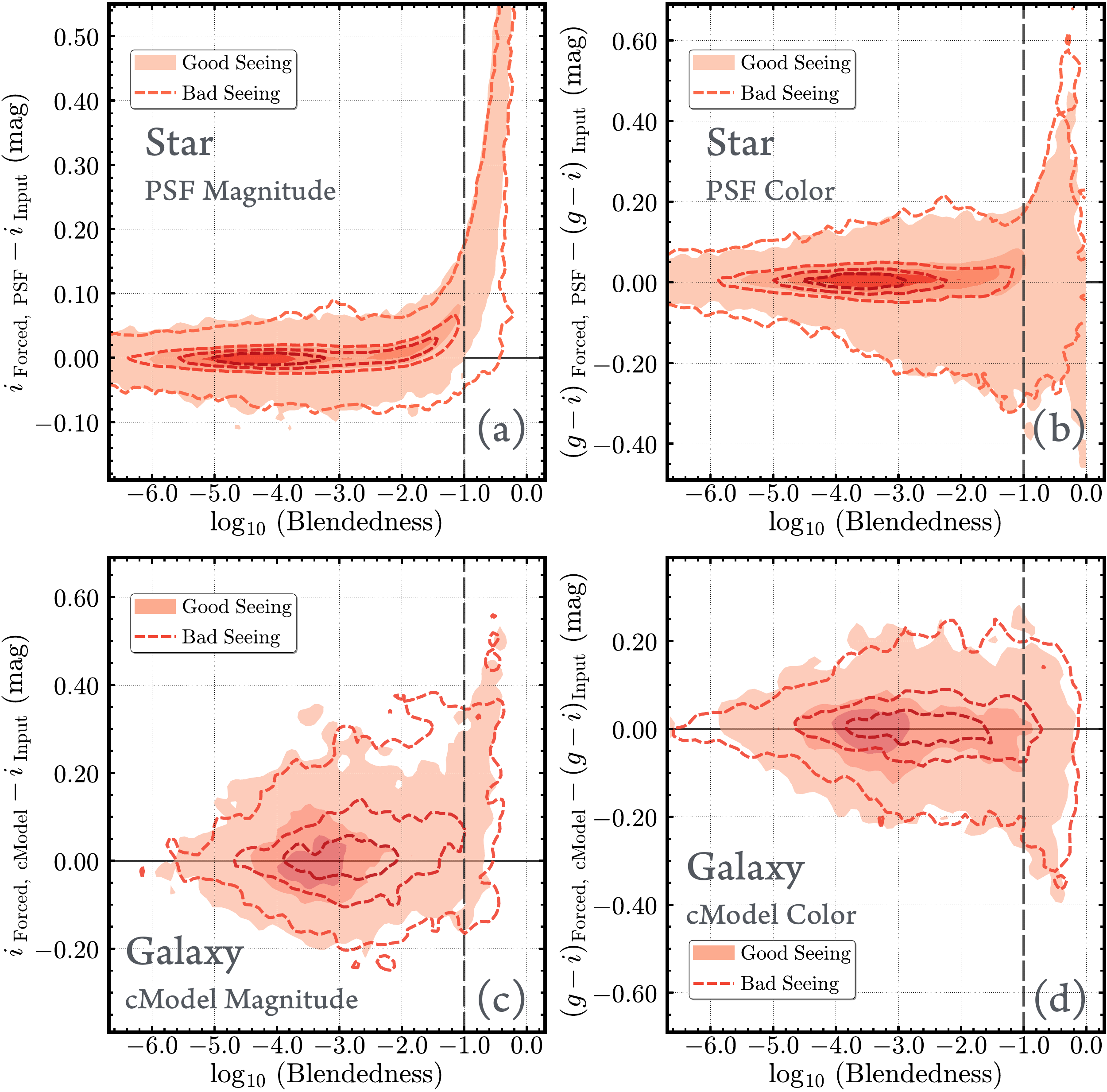}
    \end{center}
    \caption{
        Relation between the $\log_{10} (\mathrm{Blendedness})$ parameter
        and the photometric accuracy f synthetic stars (\textbf{upper row}, a and 
        b) and galaxies (\textbf{lower row}, c and d).
        \textbf{Left} columns (a and c) show the uncertainties in PSF and/or \cmodel{} 
        magnitudes. \textbf{Right} columns (b and d) show the uncertainties in 
        $(g-i)$ colors.
        Filled-contours and open-contours show results for \texttt{goodSeeing} and
        \texttt{badSeeing} \tracts{}, respectively.
        $\log_{10} (\mathrm{Blendedness}) = -1.0$ is marked using a vertical dashed 
        line.
        }
    \label{fig:blend}
\end{figure*}

\begin{figure*}
    \begin{center}
        \includegraphics[width=\textwidth]{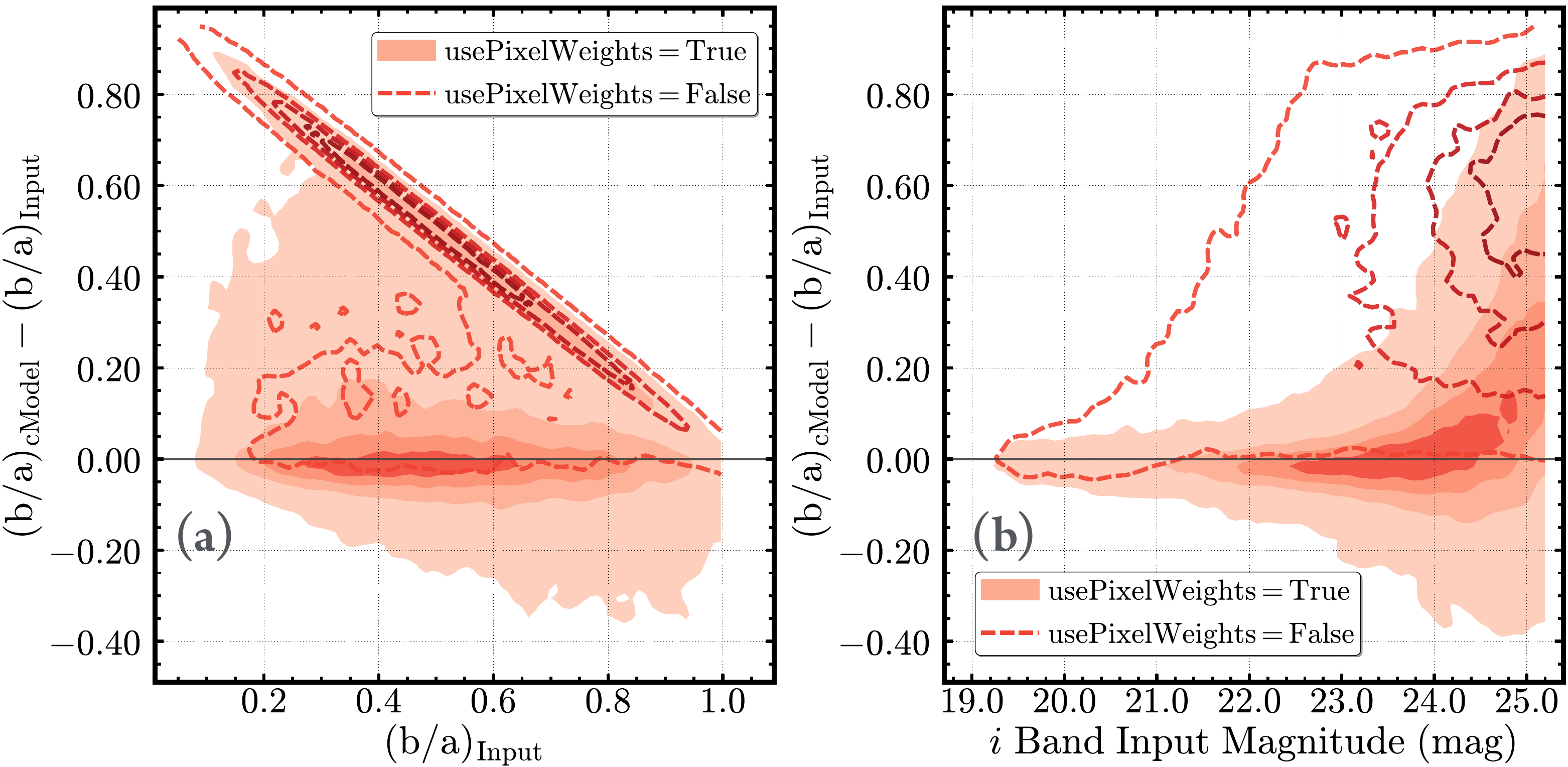}
    \end{center}
    \caption{
        Accuracy of weighted axis ratio estimates for synthetic
        galaxies using \cmodel{} photometry.
        The \textbf{left} (a) plot shows the relation between the input $(b/a)$ and 
        the uncertainty in $(b/a)$.
        The \textbf{right} (b) plot shows the relation between input magnitude and 
        the undertainty in $(b/a)$.
        Open-contours are for the default test with \texttt{usePixelWeights=False}, and 
        filled-contours correspond to \texttt{usePixelWeights=True}.
        }
    \label{fig:galaxy_ba}
\end{figure*}


\section{Other Performance Tests}
    \label{sec:others}

    In this section, we turn our attention to astrometric calibration, star-galaxy 
    separation, blending effects, and galaxy shape measurements. 

\subsection{Astrometric Calibration}
    \label{ssec:astrometry}

    We inject synthetic objects into single-\visit{} images using the initial (and 
    less accurate) astrometric calibration.
    During the image stacking step, the joint calibration process improves the
    astrometric solutions.
    Therefore, differences between the input and output coordinates from \synpipe{}
    can help us understand how much has the joint calibration step changes the 
    astrometric calibration, and how well can \hscpipe{} measure the central 
    coordinates of stars and galaxies. 

    Fig \ref{fig:astrometry} shows the distribution of astrometric offsets for 
    synthetic stars and galaxies in \texttt{goodSeeing} and \texttt{badSeeing} 
    \tracts{}.
    We find small systematic offsets at the ${\sim}10$--20 mas level, which is similar 
    to the uncertainties of astrometric calibration quoted in \citet{HSCDR1}.
    The two \tracts{} show different systematic offsets, while the synthetic stars and
    galaxies show coherent offsets in the same \tract{}.
    This shows that the initial astrometric calibration of HSC has very reasonable 
    accuracy.  
    It also suggests that \hscpipe{} can find the centroids of stars and galaxies 
    with precisions that are consistent with the astrometric calibration uncertainty
    under moderately different seeing conditions. 
    
\subsection{Star-Galaxy Separation}
    \label{ssec:sg}

    Star--galaxy separation becomes increasingly difficult at faint magnitudes 
    ($i<24.0$ mag).
    In \hscpipe{}, the star--galaxy classification primarily depends on the 
    extendedness value, which is measured by the magnitude difference between PSF and 
    \cmodel{} photometry in $i$-band.
    For more details about this algorithm, please see Bosch\etal (in prep.).

    Fig \ref{fig:sg} shows the current status of our star--galaxy classification 
    scheme.  
    Fig \ref{fig:sg} shows the difference between PSF and \cmodel{} magnitudes as a 
    function of input magnitude. 
    We find that $20$\% of synthetic stars at $i_{\mathrm{Input}}>23.0$ mag are 
    misclassified as extended objects, while only a tiny fraction of faint synthetic 
    galaxies ($i_{\mathrm{Input}}>24.0$) are misclassified as stars. 
    This confirms that the current star--galaxy classification strategy in \hscpipe{}
    results in a galaxy sample with \textbf{high completeness} and a star sample
    with \textbf{high purity}.

    It is also clear that the seeing condition strongly modifies the distributions
    of $i_{\mathrm{PSF}}-i_{\mathrm{cModel}}$. 
    Worse seeing condition makes star-galaxy separation more challenging. 
    The completeness and purity of galaxy and star samples at the faint end will 
    depend on seeing conditions.  

    In the next data release, \hscpipe{} will include an improved star--galaxy 
    classifier which will be based on a machine learning method that takes both 
    size and color information into account. 

\subsection{Blending Effects}
    \label{ssec:blendedness}

    Fig \ref{fig:blend} shows the $b$ parameter versus the difference between input and
    output values for both magnitudes and colors. 
    For both PSF and \cmodel{} magnitude distributions, we can see clear trends that 
    total fluxes are systematically underestimated for highly blended ($b>0.05$ for 
    stars or $0.1$ for galaxies) objects.
    This effect is particularly strong for stars.
    
    As for colors, higher values of $b$ results in higher statistical uncertainties 
    for both stars and galaxies, but the trends are different for each.
    Highly blended stars have clearly redder $(g-i)$ colors than the input values. 
    Highly blended galaxies seem to have $(g-i)$ colors that are bluer than the 
    true values.

    Blendedness-related uncertainties are not included in the photometric errors 
    released by \hscpipe{}.
    We therefore remind the HSC users to treat the highly blended object with
    greater caution.

\subsection{Shape and Structural Parameters of Galaxies}
    \label{ssec:shape}

    In SDSS, structural parameters like size and shape from \cmodel{} are often used to 
    study the evolution galaxies.
    Although the \cmodel{} in \hscpipe{} follows a similar algorithm, its primary
    goal now is to provide accurate and consistent magnitude and color in all five
    bands for a vast majority of faint, small galaxies.
    To improve the stability of \cmodel{} for these low-\s2n{} and barely resolved
    galaxies, \hscpipe{} uses Bayesian priors on galaxy sizes and axis ratios. 
    However, the current implementation of these priors has a serious bug and leads 
    to severe underestimated sizes and ellipticities of galaxies.
    
    Fig \ref{fig:galaxy_ba} demonstrates this issue and compares the input and 
    measured values of the \cmodel{} axis ratio. 
    Here, $({\mathrm{b}/\mathrm{a}})_{\mathrm{CModel}}$ is the flux-weighted sum of the
    axis ratios estimated by exponential and de Vaucouleurs models.    
    
    Using our \synpipe{} tests, we discovered that the bug related to the priors only 
    presents when \texttt{usePixelWeight} configuration is set to \texttt{False} during 
    the data reduction. 
    The parameter controls whether or not the per-pixel variance information is used 
    during the model fitting process.  
    As one can see, this leads to significantly overestimated axis ratios for almost
    all galaxies\footnote{It also leads to underestimated half-light radius for the same
    galaxies.}.
    We repeat the \synpipe{} run for galaxies with 
    \texttt{usePixelWeight}$=$\texttt{True}, and this change clearly mitigates
    the problem. 
    When we use \texttt{usePixelWeight}$=$\texttt{True},
    $({\mathrm{b}/\mathrm{a}})_{\mathrm{CModel}}$ provides an unbiased axis ratio 
    estimation for galaxies with $i_{\mathrm{Input}}< 24.0$ mag.
    This change is still related to the bug in priors.  
    The actual effect of using per-pixel variance should be very small.  
    The next version of \hscpipe{} will fix this bug, and \synpipe{} will help evaluate 
    the precisions of shape and size measurements. 
    This is a good example that demonstrates how the \synpipe{} framework is valuable  
    for quality control and that can be used to improve \hscpipe{}.

    For the current HSC data release, we warn users against using the sizes and shapes 
    from \cmodel{} to study the structural properties of galaxies.


\section{Summary and Conclusion}
    \label{sec:summary}

    In February 2017, the HSC survey made its first public data release, and the survey 
    continues to produce increasingly larger amounts of high-quality imaging data.
    To help achieve the designed goals of the survey and to facilitate scientific 
    investigations, we have developed \synpipe{}, a flexible framework based on
    \hscpipe{} and that can be used to perform quality control on the HSC pipeline 
    outputs. 
    \synpipe{} operates by injecting synthetic stars and galaxies into single-\visit{}
    images. 
    These data are then processed by \hscpipe{} to generate \coadd{} images and to 
    perform photometric measurements. 
    With this approach, \synpipe{} can be used as an end-to-end and realistic test of 
    the full data reduction process. 
    In this paper, we use two \tracts{} in the HSC Wide layer with representative 
    seeing conditions (\texttt{goodSeeing} with FWHM$=0.45$\asec{} seeing; 
    \texttt{badSeeing} with FWHM$=0.70$\asec{} seeing) to test the general behaviors 
    of HSC photometry for both stars and galaxies.

    Our main findings are the following:

    \begin{enumerate}

        \item The \forced{} PSF photometry provides precise and accurate 
            measurements of magnitude and color for isolated stars at 
            $18.0 < i < 25.0$.
            The typical statistical uncertainties of HSC forced PSF photometry for 
            stars ranges from 0.01 mag at $i{\sim}18.0$ mag to 0.08 mag at 
            $i{\sim}25.0$ mag (1\%-7\% precision in the $i$-band).
            The \forced{} PSF photometry can accurately recover the color--color
            sequences of stars.

        \item The \forced{} \cmodel{} photometry is reliable for synthetic galaxies
            with realistic distributions of structural parameters and colors at
            $20.0 < i < 25.0$ mag.
            The statistical uncertainties of \forced{} \cmodel{} magnitude ranges 
            from ${\sim}7$\% at $i{\sim}20.0$ mag to ${\sim}10$\% at $i{\sim}25.0$.
            The \forced{} \cmodel{} colors for galaxies are precise and are 
            consistent across in all five bands.
            They also do not show biases with input magnitudes and colors.

        \item The \forced{} PSF and \cmodel{} photometry are robust against 
            moderate changes in the seeing conditions, despite the fact that worse 
            seeing leads to a lower $\mathrm{S}/\mathrm{N}$ at fixed input magnitude.

        \item Blending ($b>0.05$) has an impact on the accuracy of both PSF and 
            \cmodel{} photometry, especially for stars.
            For highly blended stars, the \forced{} PSF photometry overestimates 
            the magnitudes of stars on average by 0.1–-0.2 mag.
            For galaxies, blending effects typically add an additional 0.05 mag
            statistical uncertainty in both magnitude and color estimates.
            
    \end{enumerate}
    
    \vspace{0.5cm}
    \noindent We also high-light several known issues with the current version 
    of \hscpipe{}:

    \begin{enumerate}

        \item The performance of the current star--galaxy separation algorithm in 
            \hscpipe{} is not perfect, and degrades at fainter magnitudes and with 
            poor seeing conditions.
            The fraction of stars that are misclassified as extended objects is still
            quite high. 
            $>20$\% of the synthetic stars at $i> 22.5$ mag are misclassified
            \footnote{Note that this does not mean high fraction of contamination 
            in the galaxy sample. Please see Bosch\etal (in prep.) for more 
            details.}

        \item The bug in priors on structural parameters and the 
            \texttt{usePixelWeights=False} nominal setting in \cmodel{} result in         
            in very biased estimates of the shapes and sizes for galaxies. 
            The axis ratio from \cmodel{} is highly overestimated and the effective 
            radius is typically underestimated.  
            This issue hence can also affect the accuracy of star--galaxy 
            classification.

    \end{enumerate}

    The HSC collaboration is working on improving these issues for future versions
    of \hscpipe{} and for subsequent data releases.  
    

    Here we focused on characterizing the photometric performance of \hscpipe{}.  
    However, \synpipe{} is a flexible tool and is being used for a range of other     
    applications. 
    For instance, Murata\etal (in prep.) uses \synpipe{} to characterize the level of
    blending in the HSC Wide Layer.  
    \citep{Niikura2017} apply \synpipe{} to high-cadence HSC observations of M31 HSC 
    to search for microlensing events. 
    \synpipe{} is also currently being used to estimate the completeness of high-$z$ 
    LBG and Lyman-$\alpha$ emitters (Ono\etal in prep.; Konno\etal in prep.), to 
    check the completeness of high-$z$ LBG pairs as a function of pair separation 
    (Harikane\etal in prep.), to test the completeness of low surface--brightness 
    dwarf galaxies selections (Greco\etal in prep.), and to test magnifications effects
    around nearby clusters (Chiu\etal in prep.).

    Both \hscpipe{} and \synpipe{} are undergoing active development. 
    Our plan is to update \synpipe{} to provide a photometric benchmark for
    each HSC survey data release, which will help improve \hscpipe{}.
    At the same time, we are also working on improving the efficiency of \synpipe{}.
    Many users are not necessarily concerned with the subtle effects involved in 
    the image stacking process. 
    Hence, a future implementation of \synpipe{} will include the option of injecting 
    objects directly onto \coadd{} images which will speed up the generation of 
    synthetic data sets. 
    The results shown in this paper demonstrate the utility of the \synpipe{} framework
    in order to perform end-to-end quality control on pipeline products to the level 
    of accuracy needed by large panchromatic surveys. 


\begin{table*}
    \begin{center}
    \begin{tabular}{| c | c | c | c | c | c | c | c | c | c | c | c |}
    \hline 
    \rowcolor[gray]{.85} \multicolumn{12}{|c|}{\large{Table}: Summary of \forced{} PSF Magnitude} \\
    \hline \hline 
    Input & Filter & \multicolumn{2}{|c|}{$g$} & \multicolumn{2}{|c|}{$r$} & \multicolumn{2}{|c|}{$i$} & \multicolumn{2}{|c|}{$z$} & \multicolumn{2}{|c|}{$y$} \\
    \cline{2-12}
    Magnitude & FWHM & $0.45$\asec{} & $0.70$\asec{} & $0.45$\asec{} & $0.70$\asec{} & $0.45$\asec{} & $0.70$\asec{} & $0.45$\asec{} & $0.70$\asec{} & $0.45$\asec{} & $0.70$\asec{} \\
    \hline
    (mag) & & (mag) & (mag) & (mag) & (mag) & (mag) & (mag) & (mag) & (mag) & (mag) & (mag) \\
    \hline
    
    \rowcolor[gray]{.85} 19.0 & \smag{} & 0.017 & 0.017  & 0.021 & 0.022 & 0.014 & 0.011 & 0.011 & 0.019 & 0.016 & 0.019 \\
    \rowcolor[gray]{.85}      & \mmag{} & 0.008 & 0.006  & 0.011 & 0.009 & 0.002 & 0.001 & 0.002 & 0.003 & 0.006 & 0.007 \\
    \hline
    
    \rowcolor[gray]{1.0} 20.0 & \smag{} & 0.018 & 0.018  & 0.023 & 0.025 & 0.015 & 0.012 & 0.012 & 0.020 & 0.017 & 0.019 \\
    \rowcolor[gray]{1.0}      & \mmag{} & 0.009 & 0.005  & 0.011 & 0.009 & 0.002 & 0.002 & 0.003 & 0.003 & 0.006 & 0.008 \\
    \hline
    
    \rowcolor[gray]{.85} 21.0 & \smag{} & 0.019 & 0.018  & 0.024 & 0.026 & 0.017 & 0.014 & 0.014 & 0.021 & 0.021 & 0.022 \\
    \rowcolor[gray]{.85}      & \mmag{} & 0.009 & 0.005  & 0.012 & 0.009 & 0.002 & 0.002 & 0.002 & 0.003 & 0.004 & 0.008 \\
    \hline 
    
    \rowcolor[gray]{1.0} 22.0 & \smag{} & 0.023 & 0.020  & 0.025 & 0.029 & 0.018 & 0.015 & 0.017 & 0.024 & 0.035 & 0.037 \\
    \rowcolor[gray]{1.0}      & \mmag{} & 0.009 & 0.005  & 0.011 & 0.009 & 0.002 & 0.002 & 0.003 & 0.004 & 0.006 & 0.006 \\
    \hline

    \rowcolor[gray]{.85} 23.0 & \smag{} & 0.031 & 0.026  & 0.029 & 0.031 & 0.021 & 0.022 & 0.025 & 0.039 &   0.076 & 0.079 \\
    \rowcolor[gray]{.85}      & \mmag{} & 0.008 & 0.004  & 0.011 & 0.009 & 0.002 & 0.002 & 0.003 & 0.005 & \n0.001 & 0.003 \\
    \hline 
    
    \rowcolor[gray]{1.0} 24.0 & \smag{} & 0.049 & 0.038  & 0.036 & 0.046 & 0.030 & 0.046 & 0.056 & 0.082 &   0.147 &   0.146 \\
    \rowcolor[gray]{1.0}      & \mmag{} & 0.007 & 0.003  & 0.010 & 0.007 & 0.002 & 0.001 & 0.005 & 0.003 & \n0.003 & \n0.033 \\
    \hline

    \rowcolor[gray]{.85} 25.0 & \smag{} &   0.119 &   0.108  & 0.087 & 0.107 & 0.062 &   0.103 &   0.124 &   0.152 & -- & -- \\
    \rowcolor[gray]{.85}      & \mmag{} & \n0.010 & \n0.008  & 0.004 & 0.006 & 0.003 & \n0.009 & \n0.001 & \n0.010 & -- & -- \\
    \hline 
        
    \end{tabular}
    \end{center}
    \caption{
        Summary of precisions and accuracies of \forced{} PSF magnitudes in all 
        five bands and in both \texttt{goodSeeing} (FWHM$=0.45$\asec{}) and
        \texttt{badSeeing} (FWHM$=0.70$\asec{}) \tracts{} based on the statistics of 
        the difference between output \forced{} PSF magnitude and input 
        value ($\Delta\mathrm{mag}$) shown in Fig \ref{fig:psf_mag}.
        Here, the precision is described using the statistical uncertainties of 
        $\Delta\mathrm{mag}$ within a series of input magnitude bins (\smag{}). 
        The accuracy of PSF photometry in the same magnitude bin is described by the 
        mean value of $\Delta\mathrm{mag}$.
        }
        \label{tab:psfmag}
\end{table*}

\begin{table}
    \begin{center}
    \begin{tabular}{| c | c | c | c | c | c | }
    \hline
    \rowcolor[gray]{.85} \multicolumn{6}{|c|}{\large{Table}: Summary of \forced{} PSF Colors} \\
    \hline \hline 
    \multirow{2}{*}{$i_{\mathrm{Input}}$} & Color & \multicolumn{2}{|c|}{$(g-i)$} & \multicolumn{2}{|c|}{$(i-y)$} \\
    \cline{2-6}
      & FWHM & $0.45$\asec{} & $0.70$\asec{} & $0.45$\asec{} & $0.70$\asec{} \\
    \hline
    (mag) & & (mag) & (mag) & (mag) & (mag) \\
    \hline
    \rowcolor[gray]{.85} 19.0 & \scolor{} &   0.023 &   0.022 &   0.018 &   0.019 \\
    \rowcolor[gray]{.85}      & \mcolor{} &   0.007 &   0.003 & \n0.004 & \n0.006 \\
    \hline 
    \rowcolor[gray]{1.0} 20.0 & \scolor{} &   0.028 &   0.024 &   0.019 &   0.019 \\
    \rowcolor[gray]{1.0}      & \mcolor{} &   0.007 &   0.003 & \n0.004 & \n0.006 \\
    \hline
    \rowcolor[gray]{.85} 21.0 & \scolor{} &   0.034 &   0.031 &   0.022 &   0.022 \\
    \rowcolor[gray]{.85}      & \mcolor{} &   0.006 &   0.002 & \n0.004 & \n0.006 \\
    \hline
    \rowcolor[gray]{1.0} 22.0 & \scolor{} &   0.053 &   0.051 &   0.031 &   0.030 \\
    \rowcolor[gray]{1.0}      & \mcolor{} &   0.004 &   0.001 & \n0.004 & \n0.005 \\
    \hline
    \rowcolor[gray]{.85} 23.0 & \scolor{} &   0.079 &   0.086 &   0.057 &   0.058 \\
    \rowcolor[gray]{.85}      & \mcolor{} & \n0.004 & \n0.015 & \n0.001 & \n0.004 \\
    \hline
    \rowcolor[gray]{1.0} 24.0 & \scolor{} &   0.116 &   0.122 &   0.111 &   0.112 \\
    \rowcolor[gray]{1.0}      & \mcolor{} & \n0.015 & \n0.001 &   0.009 &   0.004 \\
    \hline
    \rowcolor[gray]{.85} 25.0 & \scolor{} &   0.162 &   0.185 &   0.192 &   0.216 \\
    \rowcolor[gray]{.85}      & \mcolor{} & \n0.044 & \n0.018 &   0.097 &   0.097 \\
    \hline
    \end{tabular}
    \end{center}
    \caption{
        Summary of precisions and accuracies of \forced{} PSF colors using $(g-i)$ and 
        $(i-y)$ colors and in both \texttt{goodSeeing} (FWHM$=0.45$\asec{}) and
        \texttt{badSeeing} (FWHM$=0.70$\asec{}) \tracts{} based on the statistics of 
        the difference between output \forced{} PSF color and input value 
        ($\Delta\mathrm{Color}$) shown in Fig \ref{fig:psf_color}.
        Here, we describe the precision and accuracy of PSF color measurements using 
        the statistical uncertainties and mean values of $\Delta\mathrm{Color}$ in 
        bins of input $i$ band magnitudes.
    }
        \label{tab:psfcolor}
\end{table}

\begin{table*}
    \begin{center}
    \begin{tabular}{| c | c | c | c | c | c | c | c | c | c | c | c |}
    \hline
    \rowcolor[gray]{.85} \multicolumn{12}{|c|}{\large{Table}: Summary of \forced{} \cmodel{} Magnitude} \\
    \hline \hline 
    Input & Filter & \multicolumn{2}{|c|}{$g$} & \multicolumn{2}{|c|}{$r$} & \multicolumn{2}{|c|}{$i$} & \multicolumn{2}{|c|}{$z$} & \multicolumn{2}{|c|}{$y$} \\
    \cline{2-12}
    Magnitude & FWHM & $0.45$\asec{} & $0.70$\asec{} & $0.45$\asec{} & $0.70$\asec{} & $0.45$\asec{} & $0.70$\asec{} & $0.45$\asec{} & $0.70$\asec{} & $0.45$\asec{} & $0.70$\asec{} \\
    \hline
    (mag) & & (mag) & (mag) & (mag) & (mag) & (mag) & (mag) & (mag) & (mag) & (mag) & (mag) \\
    \hline
    
    \rowcolor[gray]{.85} 20.0 & \smag{} &   0.111 & 0.137  &   0.197 &   0.167 &   0.173 & 0.154 &    0.171 & 0.168 & 0.179 & 0.169 \\
    \rowcolor[gray]{.85}      & \mmag{} & \n0.006 & 0.018  & \n0.036 & \n0.024 & \n0.023 & 0.017 & \n0.006 & 0.027 & 0.011 & 0.041 \\
    \hline
    
    \rowcolor[gray]{1.0} 21.0 & \smag{} &   0.156 &   0.161  &   0.151 & 0.119 & 0.157 & 0.157 & 0.176 & 0.155 & 0.169 & 0.164 \\
    \rowcolor[gray]{1.0}      & \mmag{} & \n0.036 & \n0.008  & \n0.015 & 0.023 & 0.008 & 0.043 & 0.043 & 0.046 & 0.056 & 0.052 \\
    \hline
    
    \rowcolor[gray]{.85} 22.0 & \smag{} &   0.161 & 0.138  & 0.147 & 0.142 & 0.154 & 0.153 & 0.165 & 0.167 & 0.178 & 0.169 \\
    \rowcolor[gray]{.85}      & \mmag{} & \n0.004 & 0.029  & 0.033 & 0.050 & 0.004 & 0.040 & 0.038 & 0.045 & 0.044 & 0.048 \\
    \hline
    
    \rowcolor[gray]{1.0} 23.0 & \smag{} & 0.152 & 0.147  & 0.167 & 0.168 & 0.168 & 0.174 & 0.182 & 0.185 & 0.197 & 0.194 \\
    \rowcolor[gray]{1.0}      & \mmag{} & 0.021 & 0.040  & 0.033 & 0.036 & 0.017 & 0.021 & 0.018 & 0.028 & 0.014 & 0.020 \\
    \hline

    \rowcolor[gray]{.85} 24.0 & \smag{} & 0.176 & 0.181  & 0.179 & 0.189 &   0.183 & 0.193 &   0.201 & 0.222 &   0.241 &   0.253 \\
    \rowcolor[gray]{.85}      & \mmag{} & 0.009 & 0.028  & 0.028 & 0.025 & \n0.001 & 0.003 & \n0.002 & 0.009 & \n0.067 & \n0.057 \\
    \hline 
    
    \rowcolor[gray]{1.0} 25.0 & \smag{} &   0.196 & 0.202  & 0.211 & 0.232 &   0.218 &   0.246 &   0.241 &   0.278 &   0.308 &   0.315 \\
    \rowcolor[gray]{1.0}      & \mmag{} & \n0.005 & 0.012  & 0.012 & 0.022 & \n0.006 & \n0.013 & \n0.012 & \n0.034 & \n0.224 & \n0.299 \\
    \hline
        
    \end{tabular}

    \end{center}
    \caption{
        Summary of precisions and accuracies of \forced{} \cmodel{} magnitudes in all 
        five bands and in both \texttt{goodSeeing} (FWHM$=0.45$\asec{}) and
        \texttt{badSeeing} (FWHM$=0.70$\asec{}) \tracts{} based on the statistics of 
        the difference between output \forced{} \cmodel{} magnitude and input 
        value ($\Delta\mathrm{mag}$) shown in Fig \ref{fig:cmodel_mag}.
        Other details are the same with Table \ref{tab:psfmag}.}
        \label{tab:cmodelmag}
\end{table*}

\begin{table}
    \begin{center}
    \begin{tabular}{| c | c | c | c | c | c | }
    \hline
    \rowcolor[gray]{.85} \multicolumn{6}{|c|}{\large{Table}: Summary of \forced{} \cmodel{} Colors} \\
    \hline \hline 
    \multirow{2}{*}{$i_{\mathrm{Input}}$} & Color & \multicolumn{2}{|c|}{$(g-i)$} & \multicolumn{2}{|c|}{$(i-y)$} \\
    \cline{2-6}
      & FWHM & $0.45$\asec{} & $0.70$\asec{} & $0.45$\asec{} & $0.70$\asec{} \\
    \hline
    (mag) & & (mag) & (mag) & (mag) & (mag) \\
    \hline
    \rowcolor[gray]{.85} 20.0 & \scolor{} &   0.068 & 0.051 &   0.072 &   0.046 \\
    \rowcolor[gray]{.85}      & \mcolor{} &   0.013 & 0.002 & \n0.006 & \n0.001 \\
    \hline 
    \rowcolor[gray]{1.0} 21.0 & \scolor{} &   0.080 & 0.059 &   0.062 &   0.040 \\
    \rowcolor[gray]{1.0}      & \mcolor{} &   0.010 & 0.003 & \n0.003 & \n0.004 \\
    \hline
    \rowcolor[gray]{.85} 22.0 & \scolor{} &   0.089 & 0.072 &   0.071 &   0.059 \\
    \rowcolor[gray]{.85}      & \mcolor{} &   0.002 & 0.002 & \n0.005 & \n0.007 \\
    \hline
    \rowcolor[gray]{1.0} 23.0 & \scolor{} &   0.105 & 0.090 &   0.109 &   0.103 \\
    \rowcolor[gray]{1.0}      & \mcolor{} & \n0.018 & 0.001 & \n0.002 & \n0.005 \\
    \hline
    \rowcolor[gray]{.85} 24.0 & \scolor{} &   0.136 & 0.127 &   0.175 &   0.170 \\
    \rowcolor[gray]{.85}      & \mcolor{} & \n0.008 & 0.006 &   0.012 & \n0.007 \\
    \hline
    \rowcolor[gray]{1.0} 25.0 & \scolor{} &   0.179 & 0.185 &   0.227 &   0.241 \\
    \rowcolor[gray]{1.0}      & \mcolor{} & \n0.003 & 0.019 &   0.053 &   0.006 \\
    \hline
    \end{tabular}
    \end{center}
    \caption{
        Summary of precisions and accuracies of \forced{} \cmodel{} colors using 
        $(g-i)$ and $(i-y)$ colors and in both \texttt{goodSeeing} (FWHM$=0.45$\asec{}) 
        and \texttt{badSeeing} (FWHM$=0.70$\asec{}) \tracts{} based on the statistics of 
        the difference between output \forced{} \cmodel{} color and input value 
        ($\Delta\mathrm{Color}$) shown in Fig \ref{fig:cmodel_color}.
        Other details are the same with Table \ref{tab:psfcolor}.
    }
        \label{tab:cmodelcolor}
\end{table}


\begin{ack}
    \label{sec:ack}

    The Hyper Suprime-Cam (HSC) collaboration includes the astronomical communities of
    Japan and Taiwan, and Princeton University.
    The HSC instrumentation and software were developed by National Astronomical
    Observatory of Japan (NAOJ), Kavli Institute for the Physics and Mathematics of
    the Universe (Kavli IPMU), University of Tokyo, High Energy Accelerator
    Research Organization (KEK) in Japan,  Academia Sinica Institute for Astronomy and
    Astrophysics  (ASIAA) in Taiwan, and Princeton University in the United States.
    Funding was contributed by the FIRST program from Japanese Cabinet Office; Ministry 
    of Education, Culture, Sports, Science and Technology (MEXT); Japan
    Society for the Promotion of Science (JSPS); Japan Science and Technology Agency
    (JST); Toray Science  Foundation; NAOJ; Kavli IPMU; KEK; ASIAA; and Princeton
    University.

    This paper makes use of software developed for the Large Synoptic Survey Telescope.
    We thank the LSST Project for making their code available as free software at
    http://dm.lsstcorp.org.
    
    This work is in part supported by JSPS KAKENHI (Grant Number 26800093, 15H03654, 
    and JP17H01131) as well as MEXT Grant-in-Aid for Scientific Research on Innovative 
    Areas (No. 15H05887, 15H05892, 15H05893, 15K21733).  
    RM is supported by the US Department of Energy Early Career Award Program.
    RyM is financially supported by the University of Tokyo-Princeton strategic 
    partnership grant and Advanced Leading Graduate Course for Photon Science (ALPS).

    Pan-STARRS1 Surveys (PS1) have been made possible through contributions of
    Institute for Astronomy, University of Hawaii, Pan-STARRS Project Office,
    Max-Planck Society and its participating institutes (Max Planck Institute
    for Astronomy, Heidelberg, and Max Planck Institute for Extraterrestrial Physics,
    Garching), Johns Hopkins University, Durham University, University of
    Edinburgh, Queen's University Belfast, Harvard-Smithsonian Center for Astrophysics,
    Las Cumbres Observatory Global Telescope Network Incorporated,  National
    Central University of Taiwan, Space Telescope Science Institute,  National
    Aeronautics and Space Administration under Grant No.
    NNX08AR22G issued through Planetary Science Division of NASA Science
    Mission Directorate, National Science Foundation under Grant No. AST-1238877,
    University of Maryland, Eötvös Loránd University (ELTE), and Los Alamos
    National Laboratory.

    This research made use of
    {\texttt{Astropy}},
        a community-developed core Python package for Astronomy (\citealt{Astropy};
        \url{http://www.astropy.org/});
    {\texttt{astroML}},
        a machine learning library for astrophysics (\citealt{astroml};
        \url{http://www.astroml.org/});
    {\texttt{SciPy}},
        an open source scientific tool for Python (\citealt{SciPy};
        \url{http://www.scipy.org/});
    {\texttt{NumPy}},
        a fundamental package for scientific computing with Python (\citealt{NumPy};
        \url{http://www.numpy.org/});
    {\texttt{Matplotlib}},
        a 2-D plotting library for Python (\citealt{Matplotlib};
        \url{http://matplotlib.org/}); and
    {\texttt{scikit-learn}},
        a machine learning library in Python (\citealt{scikit-learn};
        \texttt{http://scikit-learn.org/}.

\end{ack}



\bibliographystyle{apj}



\appendix
\section{Quality Control of Synthetic Objects}
    \label{app:qc}

    We apply quality control to the stars and galaxies selected from the HSC UltraDeep
    COSMOS fields before we match them with the stars and models for galaxies from the 
    \hst/ACS COSMOS data. 
    The same quality control criteria are also used to select synthetic stars and 
    galaxies from the \synpipe{} results before we perform photometric comparisons. 
    
    To select synthetic stars and galaxies from the HSC survey data in order to test
    the photometry, we apply some basic quality control cuts.

    The HSC survey defines the full--depth \& full--color regions (\texttt{FDFC}) to
    ensure the data used for science reach the expected number of exposures in each
    band
    ($\mathtt{(\#gri\ \geq\ 4)\ and\ (\#yz\ \geq\ 6)\ and\ (Limiting\ imag> 25.6)}$;
    see \citealt{HSCDR1} for details).
    Since the two \tracts{} we used in this work are almost entirely covered in
    the \texttt{FDFC} region, we did not apply this cut.
    
    Firstly, we make sure the object is a ``primary'' detection (in the inner part 
    of \tract{} and \texttt{patch}; not a \texttt{parent} in the deblending process), 
    is successfully deblended, has reliable centroid, and is not bothered by various 
    of optical issues:

    \begin{itemize}

        \item[ ] \texttt{ detect\_is\_primary==True}
        \item[ ] \texttt{ deblend\_skipped==False }
        \item[ ] \texttt{ deblend\_nchild=0}
        \item[ ] \texttt{ [grizy]flags\_badcentroid==False }
        \item[ ] \texttt{ [grizy]centroid\_sdss\_flags==False }
        \item[ ] \texttt{ [grizy]flags\_pixel\_edge==False }
        \item[ ] \texttt{ [grizy]flags\_pixel\_interpolated\_center==False }
        \item[ ] \texttt{ [grizy]flags\_pixel\_saturated\_center==False }
        \item[ ] \texttt{ [grizy]flags\_pixel\_cr\_center==False }
        \item[ ] \texttt{ [grizy]flags\_pixel\_bad==False }
        \item[ ] \texttt{ [grizy]flags\_pixel\_suspect\_center==False }
        \item[ ] \texttt{ [grizy]flags\_pixel\_clipped\_any==False }
        \item[ ] \texttt{ [grizy]flags\_pixel\_bright\_object\_center==False }

    \end{itemize}
    
    \noindent For stars, we ensure the selected objects have useful PSF magnitude: 

    \begin{itemize}

        \item[ ] \texttt{ [grizy]flux\_psf\_flags==False }

    \end{itemize}
    
    \noindent For galaxies, we also make sure useful \cmodel{} photometry is available:

    \begin{itemize}

        \item[ ] \texttt{ [grizy]cmodel\_flux\_flags==False }

    \end{itemize}


\section{$b$: The Blendedness Parameter}
    \label{app:defineb}

    To evaluate the degree to which an object is blended with others, \hscpipe{} 
    introduces the blendedness parameter: $b$ (\texttt{[grizy]blendedness\_abs\_flux}).
    We briefly define  $b$ below; please refer to Bosch\etal (in prep.) and Murata\etal 
    (in prep.) for more details.

    For object ${\rm A}$:
    \begin{eqnarray*}
        b({\rm A}) \equiv
        1-\frac{\int_{\mathbb{R}^2} {\rm d}x~ {\rm d}y\ \mathcal{N}_{\mathrm{A}}({\bf x}\ |\  \mu_{\mathrm{A}}, {\bf\Sigma}_{\mathrm{A}})
        F_{\mathrm{A}}({\bf x})}{\int_{\mathbb{R}^2} {\rm d}x~ {\rm d}y\ \mathcal{N}_{\mathrm{A}}({\bf x}\ |\  \mu_{\mathrm{A}},
        {\bf\Sigma}_{\rm{A}}) F_{\rm{total}}({\bf x})},
        \label{eq:defineb}
    \end{eqnarray*}

    \noindent
    where $\mathcal{N}_{\rm{A}}({ \bf x }\ |\  \mu_{\rm{A}}, {\bf\Sigma}_{\rm{A}})$
    is a 2-D Gaussian function at pixel position $\bf x$, $\mu_{\mathrm{A}}$ is the
    estimated centroid of object ${\rm A}$, and covariance ${\bf \Sigma}_{\mathrm{A}}$
    is estimated based on the Gaussian-weighted adaptive second moments of object
    ${\rm A}$ (no PSF correction).
    $F_{\mathrm{total}}({\bf x })$ and $F_{\rm{A}}({\bf x })$ are pixel values of
    ${\rm A}$ before and after \hscpipe{} deblending process, respectively.

    By definition, the parameter is bound from 0 to 1.
    When the deblending process is carried out correctly, the $b({\rm A})$ parameter
    reflects the fraction of fluxes that comes from other objects in the
    region of ${\rm A}$.
    
    The current version of \hscpipe{} contains a bug related to the calculation of $b$
    parameter. 
    For bright object (e.g. $i<23.0$ mag), its effect can be ignored.  
    For fainter object, it mostly leads to underestimate of $b$ parameter by $< 12$\%, 
    especially for objects with $-3.5 < \log_{10} b < -1.5$. 
    This bug has been fixed for the future release of \hscpipe{} and HSC survey data. 
    For the \synpipe{} tests here, it does not qualitatively change the conclusions 
    related to the highly-blended objects and about the relationship between $b$ and
    photometric precision. 
    Please see Murata\etal (in prep.) and Bosch\etal (in prep.) for more details. 
    

\label{lastpage}
\end{document}